\documentclass[fleqn,usenatbib]{mnras}

\usepackage{newtxtext,newtxmath}
\usepackage{hyperref}
\usepackage{cleveref}
\usepackage[T1]{fontenc}
\usepackage{ae,aecompl}
\usepackage{graphicx}	

\usepackage{amsmath}	
\usepackage{amssymb}	
\newcommand{\ds}{\displaystyle}

\newcommand{\bs}{\boldsymbol}
\newcommand{\bm}[1]{\boldsymbol{\mathsf{#1}}}

\newcommand{\op}[1]{\boldsymbol{\mc{#1}}}

\newcommand{\vbpq} {\vec{\bf b}_{pq}}
\newcommand{\vE}{\vec{E}}
\newcommand{\mc}{\mathcal}
\newcommand{\sw}[1]{\textsc{#1}}

\DeclareMathOperator{\hard}{\op{H}}

\title[$w$-effect in interferometric imaging]{The $w$-effect in interferometric imaging: from a fast sparse measurement operator to super-resolution}

\author[A. Dabbech et al.]{
A. Dabbech,$^{1}$\thanks{E-mail: a.dabbech@hw.ac.uk}
L. Wolz,$^{2,3}$
L. Pratley,$^{4}$
J. D. McEwen$^{4}$
and Y. Wiaux$^{1}$
\\
$^{1}$Institute of Sensors, Signals and Systems, Heriot-Watt University, Edinburgh EH14 4AS, UK\\
$^{2}$School of Physics, University of Melbourne, Parkville, VIC 3010, Australia\\
$^{3}$ARC Centre of Excellence for All-Sky Astrophysics (CAASTRO)\\
$^{4}$Mullard Space Science Laboratory (MSSL), University College London (UCL), Surrey RH5 6NT, UK\\
}

\date{Accepted XXX. Received YYY; in original form ZZZ}
\pubyear{2017}

\begin{document}
\label{firstpage}
\pagerange{\pageref{firstpage}--\pageref{lastpage}}
\maketitle
\begin{abstract}
Modern radio telescopes, such as the Square Kilometre Array (SKA), will probe the radio sky over large fields-of-view, which results in large $w$-modulations of the sky image. This effect complicates the relationship between the measured visibilities and the image under scrutiny. In algorithmic terms, it gives rise to massive memory and computational time requirements. Yet, it can be a blessing in terms of reconstruction quality of the sky image. In recent years, several works have shown that large $w$-modulations promote the spread spectrum effect. Within the compressive sensing framework, this effect increases the incoherence between the sensing basis and the sparsity basis of the signal to be recovered, leading to better estimation of the sky image. In this article, we revisit the $w$-projection approach using convex optimisation in realistic settings, where the measurement operator couples the $w$-terms in Fourier and the de-gridding kernels. We provide sparse, thus fast, models of the Fourier part of the measurement operator through adaptive sparsification procedures. Consequently, memory requirements and computational cost are significantly alleviated, at the expense of introducing errors on the radio interferometric data model. We present a first investigation of the impact of the sparse variants of the measurement operator on the image reconstruction quality. We finally analyse the interesting super-resolution potential associated with the spread spectrum effect of the $w$-modulation, and showcase it through simulations. Our C++ code is available online on GitHub.
\end{abstract}


\section{Introduction}
radio interferometric imaging has been subject of numerous developments in the recent years, with the rise of the new-generation radio telescopes, namely the Square Kilometre Array (SKA)\footnote{http://www.skatelescope.org/} and its pathfinders such as the Low-Frequency Array (LoFAR)\footnote{http://www.lofar.org/}, the upgraded Karl G. Jansky Very Large Array (VLA)\footnote{https://science.nrao.edu/facilities/vla}. 
These instruments are characterised with extremely high resolution and sensitivity along with their capabilities of mapping large regions of the radio sky. To meet such specifications and hence, deliver the expected science goals, an urgent need to revolutionize radio interferometric imaging arises. Newly proposed imaging techniques in the last decade \citep{Wiaux09b,wenger10,McEwen11,Li2011,Carrillo2012,Carrillo2014,Dabbech2012,Dabbech2015, Garsden2015,onose2016,pratley16A,onose2017}, have been built almost exclusively based on sparse regularisations, and they have shown to outperform the conventional methods, namely CLEAN and its variants \citep{hogbom74,Wakker88,Cornwell2008,PratleyJ16}. These approaches solve the two-dimensional (2D) radio interferometric imaging problem, in the absence of propagation and receiver errors and within small fields of view. Under these simplified conditions, the problem consists of the recovery of the radio image from its partial and noisy Fourier samples. The imaging problem is ill-posed due to the incompleteness of the Fourier sampling, resulting from the finite number of antennas, and is challenging \textit{per se}.

In the context of wide-field radio interferometric imaging, the non-coplanarity of the radio interferometric array, its long baselines and its probed large field-of-view give rise to the so-called \textit{$w$-effect}, which originates from the third dimensionality of the Fourier space where visibilities are measured. This results in a modulation of the imaged radio sky with a chirp-like phase term parametrised by the $w$ component of the baseline in the direction of sight, also referred to as the $w$-term. Consequently, the radio interferometric imaging problem becomes three-dimensional, requiring a massive increase of memory requirements and computational cost. Yet, ignoring this effect in the imaging step can hamper significantly the quality of the reconstructed image. In the literature, several approaches have been proposed to correct for the $w$-effect, namely $w$-projection \citep{Cornwell2008} and $w$-stacking \citep{Offringa2014}. These techniques have demonstrated a reasonable compromise between memory requirements and computational cost on the one hand, and the quality of the recovered image on the other hand with respect to the traditional approaches such as faceting \citep{CornwellPerley1992}. The latter exploits the fact that the $w$ component is smaller close to the phase center. The imaged radio sky is thus decomposed into facets, where on each facet, the $w$-effect is considered flat, allowing for the use of a single point spread function (PSF) in the deconvolution step. The $w$-projection approach relies on the use of $w$-dependent convolution kernels in the Fourier domain. The data are obtained by projecting the sky Fourier transform from the plane $(u,v,w=0)$ to the correct $(u,v,w)$ plane. This implies that the two-dimensional Fourier transform can be used, hence preserving the advantages of FFTs. Whereas the $w$-stacking approach operates rather on the image domain; visibilities are gridded on discrete $w$-layers and the correction is performed on the image domain. 

Recently, $w$-projection within the context of convex optimization has been introduced \citep{Wiaux09,Wolz2013}. In \cite{Wiaux09}, the authors presented a theoretical study of the spread spectrum effect promoted by the $w$-terms within the compressive sensing theory. The latter states that the unknown signal can be accurately recovered from a few number of measurements provided that the sensing basis and the signal's sparsity basis are incoherent. In this context, the authors have presented a seminal study of how a constant $w$-modulation increases the incoherence between the sensing basis and the signal sparsity basis. As a result, the quality of the imaged radio sky is significantly enhanced. \cite{Wolz2013} demonstrated the spread spectrum effect for varying $w$ values, considering that the data arise from sparse versions of the $w$-terms in the Fourier domain, in order to lower memory requirements.
In this article, we revisit the $w$-projection for realistic settings within the same framework. We study the image reconstruction quality when the original visibilities, arising from exact non-sparse $w$-kernels, are modelled with sparse approximations of the $w$-kernels.
These models give rise to errors on the measurement operator. We provide a first study of their limits and effects on the quality of the image recovery as well as their efficiency in terms of memory requirements and accelerated computational time. Though, the general tendency is to reduce or neglect the $w$-terms due to their resulting massive computational and memory demands, we provide a proof of concept that the $w$-modulation has a large potential to promote super-resolution, recovering images beyond the resolution of the radio interferometer limited by its largest projected baseline.

The remainder of the paper is organised as follows. Section~\ref{sec:RI} describes the radio interferometric imaging problem in the presence of the $w$-term. In Section~\ref{sec:purify}, we recall the minimisation problem in radio interferometric imaging that we solve using \sw{PURIFY}, a software for radio interferometric imaging \citep{Carrillo2014,pratley16A}, into which we have incorporated $w$-projection. In Section~\ref{sec:SS}, we describe two sparsification strategies for approximating the measurement operator and also study the effect of sparsification on the image reconstruction quality. In Section~\ref{sec:SR}, we analyse and showcase the potential of the $w$-term to promote super-resolution. Finally, we state our conclusions in Section~\ref{sec:conclusions}.
\section{Radio interferometric imaging problem}
\label{sec:RI}
radio interferometers probe the electric field $\vE$ coming from the sky, each acquired measurement, called visibility, corresponds to the cross-correlation of the sensed signals by an antenna pair. Assuming a monochromatic radiation, the vectorial distance separating two antennas $(p,q)$ and denoted by $\vbpq$  is described by its coordinates in units of the wavelength $(u,v,{w})$, with $w$ is the coordinate in the direction of sight and ${\bs u}=(u,v)$ are the relative coordinates on its perpendicular plane.
The radio interferometric measurement $y({\bs u})$ is related to the sky surface brightness $\tilde{x}$ through:
\begin{equation}
y({\bs u},w)=\int {n({\bs l})^{-1}} ~a({\bs l}) ~ e^{-2i\pi {w}(n({\bs l})-1)} ~\tilde{x}({\bs l})~ e^{-2i\pi {\bs u} \cdot {\bs l}} ~{\rm d}^2{\bs l},
\label{Eq:RIP}
\end{equation}
where ${\bs l} =(l,m)$ are the coordinates of a source in a tangent plane to the celestial sphere and $n({\bs l})=\sqrt{1- \vert {\bs l} \vert^2}$ its coordinate on the line of sight, with $\vert{\bs l}\vert = \sqrt{l^2+m^2}$. We explicitly introduce the $w$-modulation term as a function $c(w,{\bs l})=e^{-2i\pi {w}(n({\bs l})-1)}$ into (\ref{Eq:RIP}). The function $ a({\bs l})$ is the illumination function of the antenna and depends on the source position. 
Other direction dependent effects (DDEs), describing unknown instrumental errors and ionospheric perturbations, can be incorporated in the same manner as the illumination function $ a({\bs l})$, even if not done so explicitly here. Note that, the term $c(w,{\bm ;})$ is also a DDE as it depends on the source position through $n({\bs l})$, yet unlike the mentioned DDEs it is known, thus \textit{easier} to model accurately although at the expense of large memory requirements. 
From now on, we call the sky intensity $x({\bs l})\equiv a({\bs l})\tilde{x}({\bs l})$; that is the modulated sky surface brightness with the illumination function. Here, we assume that the illumination function is known and invariant. However, this is not always the case in practice; $a({\bs l})$ underlies time-dependent variations, and consequently it can be described as a DDE requiring calibration. 
 
When observing a narrow field of view  $\vert{\bs l}\vert^2\ll1$ \citep[see ][]{Wiaux09} and also when the array is co-planar ${w}=0$, the ${w}$-term corresponds to a flat function $ c(w,{\bs l})=1, \forall{\bs l}$. In this case, each visibility reduces to a Fourier component of the sky intensity image. These approximations, however, break down in the case of wide-field imaging; modern radio interferometers will probe large FoVs and their antennas span hundreds and even thousands of kilometres. Futhermore, the radio sky is rather spherical \citep{McEwen08,McEwen11}. Nevertheless, to remain in the context of a 2D imaging, in the present study, we consider a relatively small field-of-view (FoV) with non-negligeable $w$-terms. Under this assumption, we adopt a first order approximation of the ${w}$-term that is a linear chirp $c(w,{\bs l})=e^{i\pi {w} {\bs l^2}}$. The $w$-term is thus considered as a norm-preserving phase modulation of the sky intensity map $x({\bs l})$ with a linear chirp of rate ${w}$. This modulation has been well studied in \cite{Wiaux09,Wolz2013} where the authors have demonstrated that the chirp modulation induces the spread spectrum effect, that is in favour of a better image reconstruction quality of the sparsity-based approaches. Note that the spread spectrum effect is not restricted to the $w$-modulation, in fact, all DDEs inherently induce similar effects due to their convolution nature in the Fourier domain. 
\subsection{The de-gridding operator}
In order to promote the use of the Fast Fourier Transform (FFT), radio measurements must lie on a regular grid (usually with a size of a power of 2). Therefore, they are modelled from uniformly sampled Fourier components via linear operations, where the \textit{continuous} (\textit{i.e.}, off the grid) measurements are interpolated using convolution kernels in the Fourier domain. In absence of the ${w}$-term and the noise, the radio interferometric measurement equation in (\ref{Eq:RIP}) can be recast in a discrete setting as follows
\begin{equation}
\label{Eq:GRIP}
{\bs y} = {{\bm G}}{\bm F}{\bm Z}{\bm D}{\bs x},
\end{equation}
where the data ${\bs y} \in \mathbb{C}^M \equiv \{y_{\ell}\equiv y({\bs u}_{\ell},w_{\ell})\}_{1\leq\ell\leq M }$ are complex measurements of size $M$. The signal to be recovered ${\bs x} \in \mathbb{R}_{+}^N$ is positive and of size $N$. The operator $\bm F$ is the Fourier transform operator. The matrix ${{\bm G}} {\in \mathbb{C}}^{(M,\alpha^2 N)}$ is sparse. Its rows ${{\bs g}}_\ell \in \mathbb{C}^{\alpha^2 N}$ are convolution kernels with $P \ll N$ non-zero elements, called de-gridding kernels, each centred around its corresponding \textit{continuous} spacial frequency point ${\bs u}_\ell=(u_\ell,v_\ell)$. The diagonal matrix ${\bm D}\in {\mathbb{R}}^{(\alpha^2 N,\alpha^2 N)}$ is correcting for the convolution in the Fourier domain; when applied to the sky, it corresponds to point-wise division of $\bs x$ with the inverse Fourier transform of the de-gridding kernel. The operator $\bm Z$ is an oversampling operator by a factor $\alpha^2$ (that is a factor $\alpha $ on each dimension often set to 2), allowing for a fine discrete uniform sampling in the Fourier domain via zero-padding of the image to be recovered. 

The de-gridding operator, not only allows for modelling the \textit{continuous} measurements, but also corrects for aliasing. In fact, it attenuates the sky-image brightness at the boundaries of the region of interest.
The choice and the support (\textit{i.e.}, number of non-zero elements) of the convolution kernels is crucial. Ideally one would use the \textit{Sinc} function for an accurate interpolation. Adopting this function translates into a multiplication of the imaged sky with a box function in the image domain, thus cancelling aliasing artefacts from sources outside the imaged region of the radio sky. However, this function is not band limited \textit{i.e.}, it has an infinite support in Fourier, thus leading to heavy computational cost comparable to the use of the Direct Fourier Transform. Therefore, band limited convolution kernels with a compact support in the Fourier domain are usually preferred. In radio interferometry, prolate-spheroidal wave functions (PSWF) are widely adopted \citep{Schwab1984,Sault1995} with a support size of $6\times6$ pixels. Kaiser-Bessel functions are also used \citep{Offringa2014} often with a support size $8\times8$ pixels. Unlike PSWFs, the Kaiser-Bessel kernels present the advantage of having an analytic expression that can be evaluated easily and accurately. Furthermore, they can be adopted with lower support size. In general, both functions lead to comparable performance \citep[see][for further discussion]{greisen98,pratley16A}. 
\subsection{The measurement operator}
\label{ss:mo}
{In this section, we introduce the $w$-term into the discrete setting of (\ref{Eq:GRIP}).} {In the image domain, the ${w}$-modulation can be expressed as a multiplication of the sky image with the ${w}$-term}. The latter can be approximated by a linear chirp ${\bs { c}_w} = \lbrace e^{i\pi {{w}} \|{\bs l}\|^2}\rbrace, \forall ~ \bs{l}$ pixel positions.
Equivalently, in the Fourier plane, the modulation corresponds to the convolution of the Fourier transform of the linear chirp with that of the signal. 
{Since the $w$-rate is baseline-dependent, each sensed measurement $y_{\ell}$ is the coefficient at the spacial frequency ${\bs u}_{\ell}$ of the signal resulting from the convolution of the Fourier transform of the linear chirp $\hat{\bs { c}}_\ell$, characterised by the rate $w_{\ell}$, with that of the sky image $\hat{\bs x}$. This means that the resulting sensed signal affected by the $w$-modulation is not the 2D conventional convolution but rather the outcome of a 2D convolution with a visibility-dependent kernel.}

For a narrow FoV denoted by $L$, the band limit of a linear chirp with a rate ${w}_{\ell}$, is deduced from its instantaneous frequency and is approximated as $B_{{w}_{\ell}} = {\vert{w}_{\ell}\vert L}/{2}$ \citep{Wiaux09}.
The largest bandwidth of the modulated signal $B'_{\rm max}$ is given by $B'_{\rm max}=\displaystyle\max_{{1\leq \ell \leq M}}{B_{w_{\ell}}}+{\parallel {\bs u_{\ell}}\parallel_2} $. Note that, in absence of $w$-modulation, the bandwidth of the signal to be recovered is $B=\displaystyle\max_{1\leq \ell \leq M}\Vert\bs u_{\ell}\Vert_2$, which constitutes the band limit of the radio interferometer. The true radio sky is characterized with an infinite bandwidth. Therefore, a correct modelling of the visibilities requires accommodating the bandwidth of the unknown signal, at least, up to the largest bandwidth $B'_{\rm max}$, that is accessible by the highest $w$-modulation. This translates in choosing a pixel-size $\Delta l' = 1/2 B'_{\rm max} $ in the image domain, that is smaller than the radio interferometer resolution (\textit{i.e.} $\Delta l =1/ 2B$). Consequently, the imaging step consists in estimating the unknown signal at a resolution above the instrument band limit. Yet, in practice a convolution of the deconvolved image of the sky with a low-pass filter, that is smoothing kernel corresponding to the instrument resolution, is often  applied. Hence, all the spatial Fourier modes above the instrument's band limit are set to zero.
When accounting for the $w$-term, each measurement $y_\ell$ reads:
\begin{equation}
\label{eq:kernel_conv}
y_\ell = \tilde{\bs g}_\ell^\top ~{\bm F}{\bm Z}{\bm D}\bs{x} \text{, with } \tilde{\bs g}_\ell ={\hat{\bs{c}}}_\ell\ast{{\bs g}}_\ell,
\end{equation}
where $\ast$ denotes linear convolution.
The estimated image $\bs x$ is of size $N'$, with a bandwidth $B'_{\rm max}$.
The vector ${\hat{\bs{c}}}_\ell \in \mathbb{C}^{{N}''}$ denotes the Fourier transform of the chirp modulation of size ${N}''=\alpha^2N'$ and support $K_{{w}_{\ell}}<N''$. $\tilde{\bs{g}}_{\ell} \in \mathbb{C}^{{N}''} $ are convolution kernels coupling the de-gridding interpolation kernels and the ${w}_{\ell}$-terms, with a support $N_{\ell} = P+K_{{w}_{\ell}}$. 
The radio measurements are hence modelled as:
\begin{equation}
\label{eq:ra-problem0}
{\bs y} = \tilde{\bm{G}} {{\bm F}{\bm Z}{\bm D}} {\bs x} +{\bs n} = {\bs {\Phi}} \bs x +\bs n,
\end{equation}
where $\tilde{\bm{G}}$ is the $w$-projection operator, whose rows correspond to ${\tilde{\bs g}}_{1 \leq \ell\leq M}\in \mathbb{C}^{N''}$, and ${\bs n}\in \mathbb{C}^M$ is additive white Gaussian noise, with known statistics, modelling the instrumental noise. The operator ${\bs\Phi} \equiv{ \tilde{\bm {G}}} {{\bm F}{\bm Z}{\bm D}} $ is the measurement operator, incorporating the ${w}$-modulation. The problem of recovering the signal $\bs x$ from the measurements $\bs y$ is ill-posed due to the incomplete sampling and the noise \textit{i.e,} it does not present a unique solution. Therefore, additional prior information on the signal to be recovered $\bs x$ is usually imposed to better constrain the imaging problem and consequently obtaining a good approximate of the unknown image.
\section{Sparse image reconstruction}
\label{sec:purify}
Numerous works have demonstrated the applicability of compressive sensing theory to solve the radio interferometric imaging problem. Within this framework, the unknown signal is assumed to be sparse or compressible in a data representation space ${\bs {\Psi}}$ (\textit{e.g.} wavelets, curvelets, discrete cosine transform). This is generally the case for natural signals. The theory of compressive sensing states that the exact recovery of the unknown signal is possible from a small number of measurements $M$ that is below the Shannon-Nyquist sampling rate, provided that the sensing operator $\bs {\Phi}$ and the signal's sparsity basis ${\bs {\Psi}}$ are incoherent \citep{Candes2006,donoho06}; the mutual coherence defined as the largest cross-correlation between the columns of ${\bs {\Psi}}$ and ${\bs {\Phi}}$.

In radio interferometry, the Fourier basis constitutes the sensing basis. Furthermore, assuming sparsity-by-synthesis of the signal $\bs x$ in a dictionary $\bs {\Psi}$, that is $\bs x = \bs{ \Psi} \pmb {\alpha}$, where the synthesis coefficients vector $\pmb \alpha$ is sparse, the mutual coherence of the sparsity and the sensing basis is defined as the largest modulus of the Fourier coefficients of the basis functions constituting the sparsity basis $\bs \Psi$. For instance, considering a signal of size $N$, sparse in its domain, its sparsity basis $\bs \Psi$ is the Dirac basis and the mutual coherence of the Fourier basis and the Dirac basis is given by $1/ \sqrt N $. For large $N$, the mutual coherence goes to zero. Therefore, the Dirac basis is totally incoherent with the Fourier basis. For more complex signals, sparsity is assumed in more sophisticated dictionaries such as wavelets. The latter can be sufficiently incoherent with the Fourier basis.  Note that, the choice of the adequate sparsity basis is important and is application-dependent. In radio astronomy, several dictionaries have been adopted and have shown to be relevant, namely wavelets such as the isotropic undecimated wavelet transform adopted in \cite{Li2011,Dabbech2012,Dabbech2015,Garsden2015}, the Daubechies wavelets concatenated with the Dirac basis in \cite{Carrillo2012,Wolz2013,Carrillo2013,Carrillo2014,onose2016,pratley16A,onose2017}. 

The $w$-term, as a norm-preserving modulation, spreads the energy of the Fourier coefficients of the sparsity dictionary's basis functions over the neighbouring Fourier modes, reducing the amplitudes of these Fourier coefficients. Consequently, the mutual coherence of the sparsity dictionary and the sensing basis is significantly decreased. Therefore, intrinsically, the $w$-term reinforces the embedding of the radio interferometric imaging problem within the compressive sensing theory, a detailed theoretical study on the matter is provided in \cite{Wiaux09}.
\subsection{Imaging with convex optimisation}
To solve the inverse problem given in (\ref{eq:ra-problem0}), we adopt sparsity-by-analysis promoting prior as a reguliser. This imposes that the projection of the unknown signal in a sparsity basis $\bs \Psi$ is sparse, that is $\pmb \alpha = \bs{\Psi}^\dagger \bs x$ and the analysis vector $\pmb \alpha$ being sparse. The adopted sparsity basis is that proposed in \cite{Carrillo2012}, $\bs \Psi$ is over-complete and is chosen to be the concatenation of nine basis; the Dirac basis and the eight Daubechies wavelet basis. The minimisation problem reads:
\begin{equation}
\label{eq:min-pb}
\displaystyle \min_{\bs x} \parallel {\bs \Psi}^\dagger {\bs x}\parallel_{1} \text{ s.t. } \left\{
\begin{array}{l}
 \parallel{\bs y}-{\bs \Phi}{\bs x}\parallel_2 \leq \epsilon, \\
 {\bs x\geq \bm 0},
\end{array}
\right.
\end{equation}
where the bound $\epsilon$ is the $\ell_2$ norm of the noise $\bs n$. To impose sparsity on $\bs{\Psi}^\dagger \bs x$, intuitively one should minimise the $\ell_0$ norm  of  $\bs{\Psi}^\dagger \bs x$, however this leads to NP-hard problem. Therefore, the $\ell_0$ norm is often relaxed by adopting the $\ell_1$ norm, that is sparsity-promoting and convex. Re-weighted schemes of the $\ell_1$ norm have been proposed in the literature to better approximate the $\ell_0$ norm \citep{Candes2008}. Furthermore, they have shown considerable improvements of the image recovery quality \citep{Carrillo2012,Carrillo2014}. 

The constrained minimisation problem (\ref{eq:min-pb}) can be efficiently solved using convex optimisation, furthermore, it can be written as follows
\begin{equation}
\label{eq:min-pb-cx}
\displaystyle \min_{\bs x} \parallel {\bs \Psi}^\dagger {\bs x}\parallel_{1} +\iota_{\mc{\mathbb{R_+^{N}}}}(\bs{x}) +\iota_{\mc{B}}(\bs{\Phi \bs x}),
\end{equation}
where, the constraints defined in (\ref{eq:min-pb}) are imposed using the indicator function $\iota_{\mc{C}}$, defined on a non-empty set ${\mc{C}} $ as
\begin{equation}
	\iota_{\mc{C}} (\bs{z}) \overset{\Delta}{=} \left\{ \begin{aligned}
					0 & \qquad \bs{z} \in \mc{C} \\
					+\infty & \qquad \bs{z} \notin \mc{C}
				 \end{aligned} \right. 
	\label{indicator-function}
\end{equation}
Minimising the function $\iota_{\mc{\mathbb{R_+^{N}}}} $ imposes the positivity and reality constraints of the signal to be recovered $\bs x$. The function $\iota_{\mc{B}}$ applied to $\bs \Phi \bs x$ is the data fidelity term. When minimised, it constrains the residual to be in the $\ell_2$ ball $\mc B$. The latter is a convex set, characterised by the noise level $\epsilon$ such that
 $\mc{B} = \{ \bs{z} \in \mathbb{C}^M: \| \bs{z} - \bs{y} \|_2 \leq \epsilon \}$.

The problem (\ref{eq:min-pb-cx}) fits well within proximal splitting methods. These are solving for optimisation problems of the form
\begin{equation}
\displaystyle \min_{\bs x} g_1(\bs x)+ g_2(\bs x) \dots +g_n(\bs x),
\end{equation} 
where the functions $g_{i \in \{1\dots n\}}$ are proper, convex and lower semi-continuous. Each function is minimised individually. Solutions to differentiable functions, such as the $\ell_2$ norm, are obtained using their gradient. While, minimising non-smooth functions involve the use of their proximity operators, in particular the $\ell_1$ norm is minimised through its proximal operator, that is the soft-thresholding. We direct the reader to \cite{combettes09} for a good review. These methods have gained wide interest in the recent years thanks to the splitting of the functions, hence their scalability to large-scale problems. In particular, several proximal splitting methods have been proposed to solve (\ref{eq:min-pb-cx}) for radio interferometry. \cite{Carrillo2012} used the Douglas-Rachford splitting method \citep{combettes07}, \cite{Carrillo2014} adopted the simultaneous-direction method of multipliers (SDMM) \citep{Setzer2010}. More recently, \cite{onose2016} have proposed two highly parallelisable and distributed algorithmic structures based on the alternating direction method of multipliers (ADMM) \citep{Boyd2011} and the Primal-Dual (PD) algorithm \citep{Komodakis2015}, allowing for full splitting of the data and the linear operator into blocks.
\subsection{Radio interferometric imaging in \sw{PURIFY}}
In our study, we use the software PURIFY\footnote{http://basp-group.github.io/purify/} \citep{Carrillo2014,pratley16A}, that is a package written in C++ for radio interferometric imaging and using $\ell_1$ regularisation-based solvers, in particular, the SARA algorithm originally proposed in \cite{Carrillo2012}. \sw{PURIFY} supports standard de-gridding kernels \citep{pratley16A} in particular Kaiser-Bessel convolution kernels, with optimized parameters suggested in \cite{Fessler2003} and the prolate spheroidal wave kernels as described in \cite{Schwab1984}. We have implemented the $w$-terms as described in (\ref{eq:kernel_conv}) leading to the $w$-projection operator $\tilde{\bm G}$ detailed in Section \ref{sec:RI}. Variants of the measurement operator are also included where sparse models of the $w$-projection operator $\tilde{\bm G}$ through adaptive sparsification strategies of its convolution kernels are adopted, these are detailed in Section~\ref{sec:SS}. 

\sw{PURIFY} calls the Sparse OPTimisation (SOPT) software package\footnote{http://basp-group.github.io/sopt/}, a collection of solvers of the convex problem (\ref{eq:min-pb-cx}), namely the simultaneous-direction method of multipliers (SDMM) algorithm proposed by  \cite{Carrillo2014} and the proximal alternating direction method of multipliers  proposed by \cite{onose2016}. The PD-based splitting algorithm proposed in the latter article will be included in a future release of SOPT. While SDMM is known to be computationally heavy since it involves matrix inversion, both algorithmic structures proposed in \cite{onose2016} are parallelisable and allow for an efficient distributed implementation, through the splitting of the data into blocks\footnote{In particular, PD-based splitting algorithm has shown to be very flexible allowing for full data and operators splitting, resulting in lower computational burden and memory requirements. It is also prone to further increased scalability by using randomized updates; where only randomly selected data blocks are processed within each iteration.}. In the present study, we have adopted the ADMM algorithm as a solver. Note that, the block splitting feature will be included in a future release of PURIFY.

\section{Sparse representations of the measurement operator}
\label{sec:SS}
{ The $w$-projection approach allows for an accurate correction of the $w$-term, accounting for the exact $w$-modulation through its discrete Fourier transform. The resulting $w$-terms, however, are usually non-sparse resulting in a high-dimensional, non-sparse measurement operator.}
In fact, in the presence of high $w$-rates their corresponding chirp kernels $\hat {\bs c}$ do not decay rapidly to zero. Moreover, even in the presence of small $w$-rates, their corresponding $\hat{\bs c}$ can present large amounts of very small coefficients yet non-zero valued, and so they are rather compressible but not strictly sparse, the same applies for the resulting $w$-projection operator $\tilde{\bm G}$. This is computationally demanding as it involves a heavy measurement operator ${\bs {\Phi}}$. In practice, \cite{Cornwell2008} suggest to consider truncated $w$-terms such that coefficients below $10\%$ of the maximum coefficient in terms of modulus are set to zero. Moreover, the $w$-rates are further sampled on a discrete small set of values. These considerations result in faster operators, though they might introduce large errors on the model of the measurement operator, hence hampering the reconstruction quality and the dynamic range in particular. 

We shall define the Fourier-transformed Chirp operator $\hat{\bm C}$ whose rows ${\hat{\bs c}}_{1\leq \ell \leq M}\in \mathbb{C}^{N''}$ are the Fourier transforms of the linear chirp modulations with the rates $w_{{1\leq \ell \leq M}}$, centred at the zero frequency.
In the context of discrete measurements on a uniformly sampled grid, no de-gridding kernels are involved and the $w$-projection operator is given by the Fourier transformed $w$-terms ${\hat{\bs c}}_{\ell}$, each centred at its corresponding $\bs u_{\ell}$ point. In this case, \cite{Wolz2013} show that the $w$-projection operator, when expressed as a sparse matrix, results in low effective computational time and memory requirements for imaging, provided that it is remains accurate.

In this study, we present two approaches for sparse representations of the measurement operator. The first strategy is based on the sparsification of the Fourier-transformed Chirp operator $\hat{\bm C}$ before the convolution with the de-gridding operator ${\bm G} $ and resulting in a sparse $w$-projection operator $\tilde{\bm G}$. The second approach consists in the direct sparsification of the $w$-projection operator $\tilde{\bm G}$. In the first approach, thanks to the reduced number of the non-zero elements of $\hat{\bm C}$, the process of the row-wise convolution of the de-gridding operator ${\bm G}$ and $ \hat {\bm C}$ is accelerated since the number of non-zero elements in each $w$-term is significantly reduced.  This is not the case in the direct sparsification of the $w$-projection operator $\tilde{\bm G}$, since the convolution is performed using the non sparse $w$-terms, though in general this approach yields more sparsity of the measurement operator as it is shown in this section.

To determine a sparse representation of the $w$-projection operator $\tilde{\bm G}$, we adopt the sparsification technique proposed in \cite{Wolz2013}. The idea is to apply an adaptive hard-thresholding on each row ${\bs r}_{\ell}$ of an operator $\bm R \in \mathbb{C}^{M\times N}$, that is a component-wise operation. The hard-thresholding operator denoted by $ \hard$ is defined as follows
\begin{equation}
	\Big( \hard_{\tau_{\ell}}(\bs{r}_{\ell}) \Big)_i \overset{\Delta}{=} \left\{ 
	\begin{array}{cl}
		\ds r_{i,\ell} & \qquad | r_{i,\ell} | > \tau_{\ell}\\
		\ds 0 & \qquad | r_{i,\ell} | \leq \tau_{\ell}\\
	\end{array}\right. \quad \forall i,
	\label{prox-L1}
\end{equation}
 with the constraint of loosing a fixed energy fraction $\gamma$ across all the rows. The energy of a vector $\bs r_{\ell} \in \mathbb{C}^{M}$ is defined as the sum of the squared modulus of its elements $E^{\rm{total}}_{\ell}\equiv\sum_{i=1}^N |r_{\ell,i}|^2$. For a preserved energy $E^{\rm{sparse}}_{\ell}= (1-\gamma) \times E^{\rm{total}}_{\ell}$, where $0\leq \gamma \leq 1$, the significant non-zero elements of $\bs r_{\ell}$ are determined by computing a threshold $\tau_{\ell}$ using a bisecting method such that elements contributing to the energy $E^{\rm{lost}}_{\ell} = E^{\rm{total}}_{\ell} - E^{\rm{sparse}}_{\ell} $ are set to zero. This approach allows to adaptively reduce the number of significant elements for any given operator while not limiting the process to a specific support size. A sparse approximation of the $w$-projection operator $\tilde{\bm G}$ is obtained by applying 
this sparsification technique on the rows ${\hat{\bs{c}}}_{{1\leq \ell \leq M}}$ of $\hat{\bm C}$ or the rows $\tilde{\bs{g}}_{1 \leq \ell \leq M}$ of $\tilde{\bm G}$. 

The present study consists in studying the effect of adopting sparse approximations of the $w$-projection operator in the image recovery step. This results in model errors of the radio interferometric measurements. We therefore provide limit values of the sparsification levels considered in the sparse approximations of the $w$-projection operator, for additive noise $\bs n$ characterised by two input signal-to-noise ratios. In these limit cases, degradation of the image reconstruction quality is observed;  the errors of the adopted backward model are no longer buried in the noise. We shall recall that in \cite{Wolz2013}, the authors have shown that sparse Chirp operators do enhance the image reconstruction quality. Exact sparse $w$-projection operators have been employed for imaging; in both the forward model, adopted for radio interferometric data simulation following (\ref{eq:ra-problem}), and the backward model, considered in the image recovery using the approach described in (\ref{eq:min-pb}). 
\subsection{Simulations settings}
\begin{figure}
\centering
\includegraphics[scale=0.25]{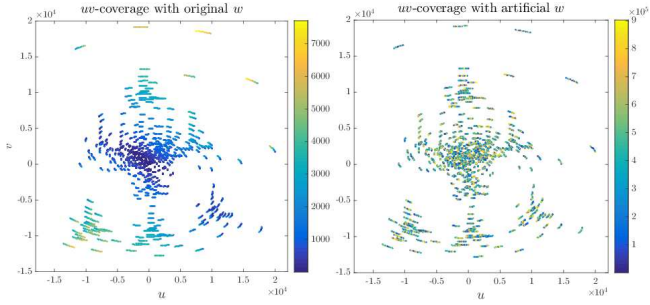}
\caption{{\small{Adopted ASKAP $uv$-coverage for the study of sparse approximations of the measurement operator, with $M =7560$ measurements. The $u$ and $v$ components correspond to $x$-axis and $y$-axis respectively, colors correspond to the $w$ components in absolute values. Left: \textit{naturally} generated $w$ components, right uniformly distributed $w$ components for $w_f=0.1$. }}}
\label{fig_wrates}
\end{figure}

We study the explained strategies for a sparse measurement operator through realistic simulations of radio interferometric data. A realistic $uv$-coverage is simulated using  the Meqtrees software \citep{Noordam2010}  from the antenna configuration of  the Australian Square Kilometre Array Pathfinder (ASKAP)\footnote{http://www.atnf.csiro.au/projects/askap/index.html} (see Fig.~\ref{fig_wrates}, left panel). The coverage is obtained for a total observation time of 1 hour and time spacing $\delta{\rm t}=5$ minutes, pointing at declination $-10d0m0s$ and right-ascension $0d0m0s$. The total number of $\bs u$ points is $M=7560$. The $w$ components generated in this setting are extremely small leading to chirp kernels that can be easily approximated by a Dirac for nearly $90\%$ of the $uv$-coverage. Therefore, in our tests, we consider artificially large $w$ values, which we generate as zero-mean uniform distribution $w$ values - hence not correlated to the $uv$ points distribution -, such that the maximum $w$-rate $|w_{\rm max}|=w_f \times w^*$, where $w^* =2B/{L^2}$ is the $w$-rate resulting in a $w$-modulation having the bandwidth $B$ of the signal. Hence, the parameter $w_f$ translates the fraction of the maximum $w$-modulation with respect to $w^*$. In Fig.~\ref{fig_wrates}, we showcase an example of uniformly distributed $w$ values for $w_f=0.1$ versus the naturally generated $w$ values, where the maximum $w$-rate corresponds to $w_f = 8.5\times 10^{-4}$.

\begin{figure}
\centering
\includegraphics[scale=0.27]{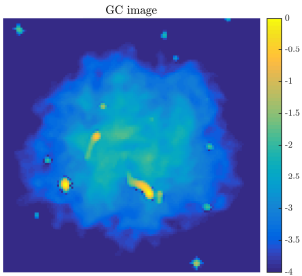}
\caption{{\small{ Galaxy cluster image $\rm{GC}$ of size $128\times128$ pixels, displayed in $\rm{log}10$ scale. }}}
\label{fig_sparse_GC}
\end{figure}

The considered ground-truth sky image is an image of a galaxy cluster ($\rm{GC}$), obtained using the FARADAY tool \citep{murgia2004} (see Fig.~\ref{fig_sparse_GC}, right panel). The $\rm{GC}$ image is a collection of very bright compact sources corresponding to galaxies and a central diffuse faint emission that corresponds to a radio halo. Beside its diversity as an image, in terms of structure, it is also characterized with a high dynamic range of order $10^6$. We consider a low resolution version of the GC image, that is of size $N=128\times128$ pixels with its highest pixel value scaled to 1 and its bandwidth is equal to the radio interferometric band limit, that is $B=\displaystyle\max_{1\leq \ell \leq M}\parallel {\bs u_{\ell}}\parallel_2$. The observed field of view is narrow with $L=0.21$ degrees and is determined by the simulated baseline coverage and the size of the ground-truth image. We generate radio interferometric visibilities at the frequency 1GHz, in \sw{MATLAB}. The measurements are corrupted with additive white Gaussian noise, with input signal-to-noise ratio (iSNR) of 30dB and 40dB. For the de-gridding kernels, we adopt the Kaiser-Bessel interpolation kernels and fix their support size to $4\times4$ pixels. Note that, to accommodate the full $w$-modulation kernels up to the bandwidth $B'_{\rm max}$ as described in subsection \ref{ss:mo}, we apply an up-\textit{sampling} operator to the sky image, which consists in performing a zero-padding in the Fourier domain. Simulated observations are thus given by:
\begin{equation}
\label{eq:ra-problem}
{\bs y} = \tilde{\bm{G}} {{\bm F}{\bm Z}{\bm D}{\bm U}} {\bs x} +{\bs n} = {\bs {\Phi}} \bs x +\bs n,
\end{equation}
where $\bm U$ is the up-sampling operator such that the up-sampled image $\bm U \bs x$ is of size $N'$ (determined with respect to the highest $w$-modulation considered, see subsection \ref{ss:mo}). The measurement operator is 
$\bs \Phi = \tilde{\bm{G}} {\bm F}{\bm Z}{\bm D}{\bm U}$, and its adjoint operator is
${\bs \Phi}^{\dagger} = \tilde{\bm U} {\bm D} \tilde{\bm Z} {\bm F^{\dagger}} {\tilde{\bm{G}}^{\dagger}} $, where the operator $\tilde{\bm Z} $ is undoing the zero-padding in the image space and $\tilde{\bm U}$ is the down-sampling operator.

For image recovery, we adopt the ADMM algorithm solving for the minimisation problem (\ref{eq:min-pb}). For further details on the ADMM algorithm, we direct the reader to the work of \cite{onose2016}. We consider the following parameters; the parameter controlling the proximal operator of the $\ell_1 $ norm is set to $\kappa = 10^{-3}$ and the stopping criteria which are the maximum number of iterations $I_{\rm max} =2500$, the variation on the solution $\beta=5\times10^{-6}$. The bound $\epsilon$ on the ${\ell}_2$-ball is set from the $\chi^2$ distribution of the noise as $\epsilon ^2 = (2M + 4 \sqrt M) \sigma^2 /2$, where $\sigma^2 /2$ is the variance of both the real and imaginary part of the noise \citep{Carrillo2012,onose2016} and on which we allow an upper bound that is $\epsilon' =1.005\times \epsilon$. To assess the reconstruction quality, we adopt two metrics. These are the signal-to-noise ratio (SNR) and the minimum ratio metric (MR), defined as 
\begin{align*}
{\rm SNR} = 20 \log_{10} \frac{|| {\bs x}||_2}{|| {\bs x}-{\bar{\bs x} }||_2}, 
\end{align*}
and
\begin{align*}
{\rm MR}=\frac{1}{N} \sum_{i=1}^{N} \min( \frac{x_i}{{\bar x}_i}, \frac{{\bar x}_i}{x_i}),
\end{align*}
where $\bar{\bs x}$ denotes the estimated signal of the ground-truth signal $\bs x$. 
Compared to the SNR, the MR metric can be more sensitive to pixel-wise errors, in particular for high dynamic range images such as the $\rm{GC}$ image, where the $\rm{SNR}$ metric tends to be dominated by the brightest structures, containing the highest energy of the image. 

\subsection{Sparsification of the Fourier-transformed Chirp operator}
We study the sparsification technique to each row $\hat{\bs{c}}_\ell$ of the operator $\hat{\bm C}$ preserving the energy $E^{\rm sparse}_{\ell} = (1-{10}^{n_{\rm sparse}})\times E^{\rm{total}}_{\ell} $ with $n_{\rm sparse} \in \{-6, -5, -4.5, -4, -3.5, -3, -2\}$ denoting the sparsification level. Since the chirp kernel is a norm-preserving modulation, $E^{\rm{total}}_{\ell}=1$, $\forall \ell$. We consider various $w$-rates within 5 different ranges, such that $ \forall ~ {1\leq \ell \leq M},~|w_{\ell}|\leqslant w_f \times w^*$, with $w_f\in\{0.1, 0.2, 0.3,0.4, 0.5\}$.
{To quantify the sparsity of an operator, we adopt the sparsity ratio (SR) metric that is defined as the ratio between the cardinality of an operator $\bm R$ and the total number of its elements; ${\rm SR}={\rm{card}(\bm R)}/({N'' \times M})$.
\begin{figure}
	\centering
	\begin{minipage}{.99\linewidth}
	\centering
	\includegraphics[width=0.75\textwidth, height=0.4\textwidth]{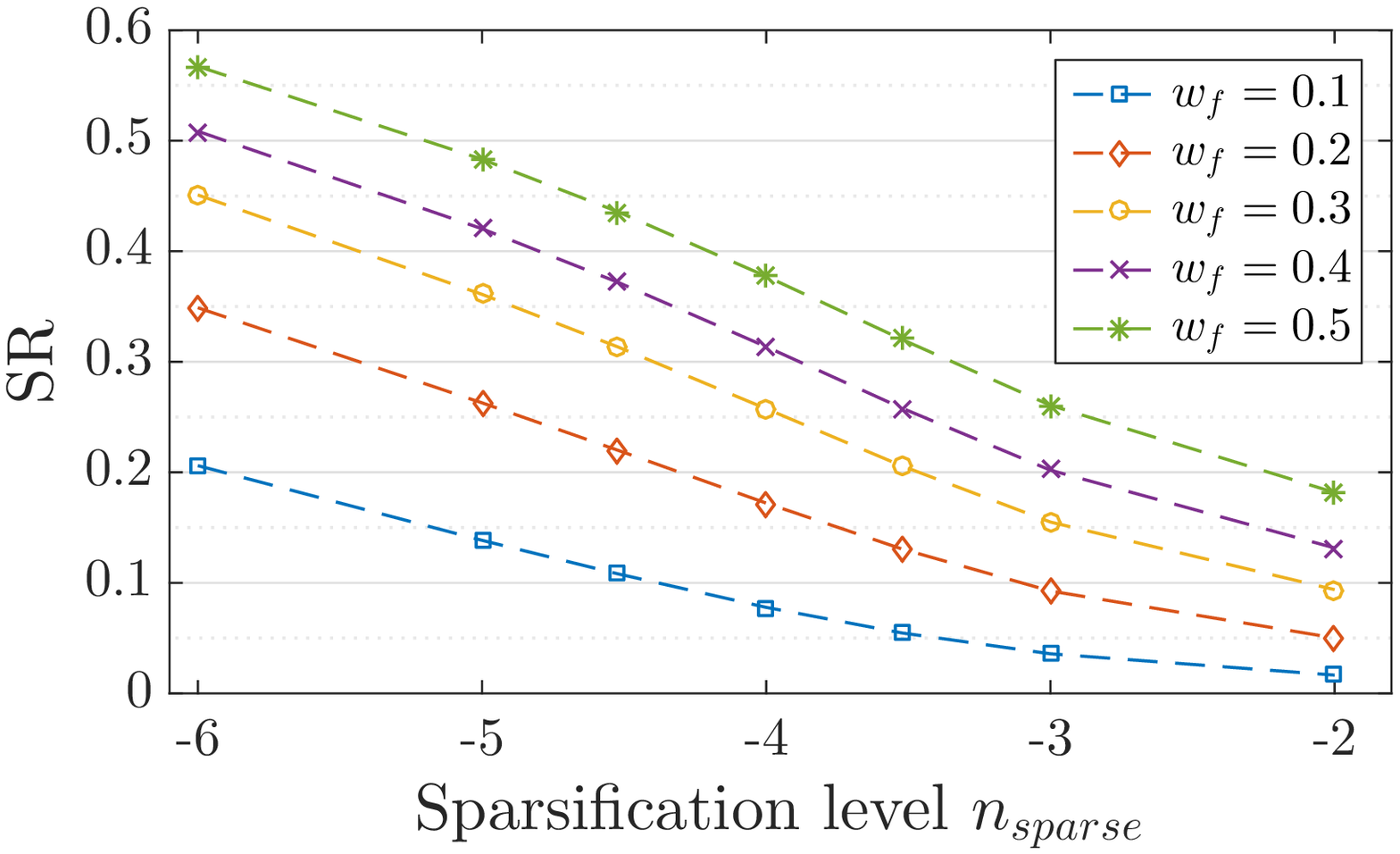}
 \caption{\small Sparsity ratios ($y$-axis) of the Fourier-transformed Chirp operator $\hat{\bm C}$. Sparsification is applied on the operator $\hat{\bm C}$ using different sparsification levels $n_{\rm sparse}$ ($x$-axis). }
 \label{fig_Csparsity}
 \end{minipage}	
	\begin{minipage}{0.99\linewidth}
	\vspace{3mm}
	\centering
 \includegraphics[width=0.85\textwidth]{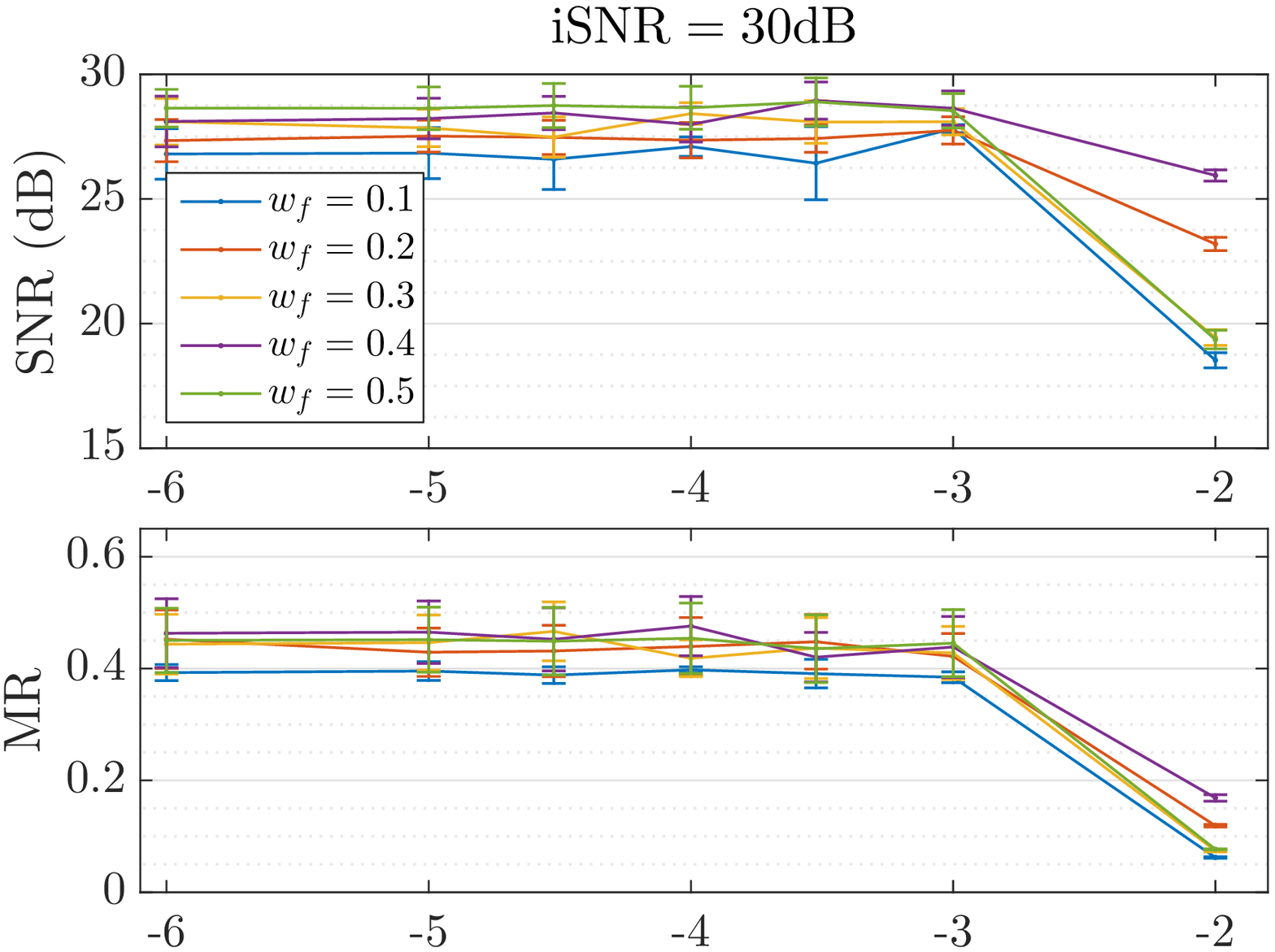}
 \vspace{3mm}
	\end{minipage}
  \begin{minipage}{0.99\linewidth}		
	\centering
 \includegraphics[width=0.85\textwidth]{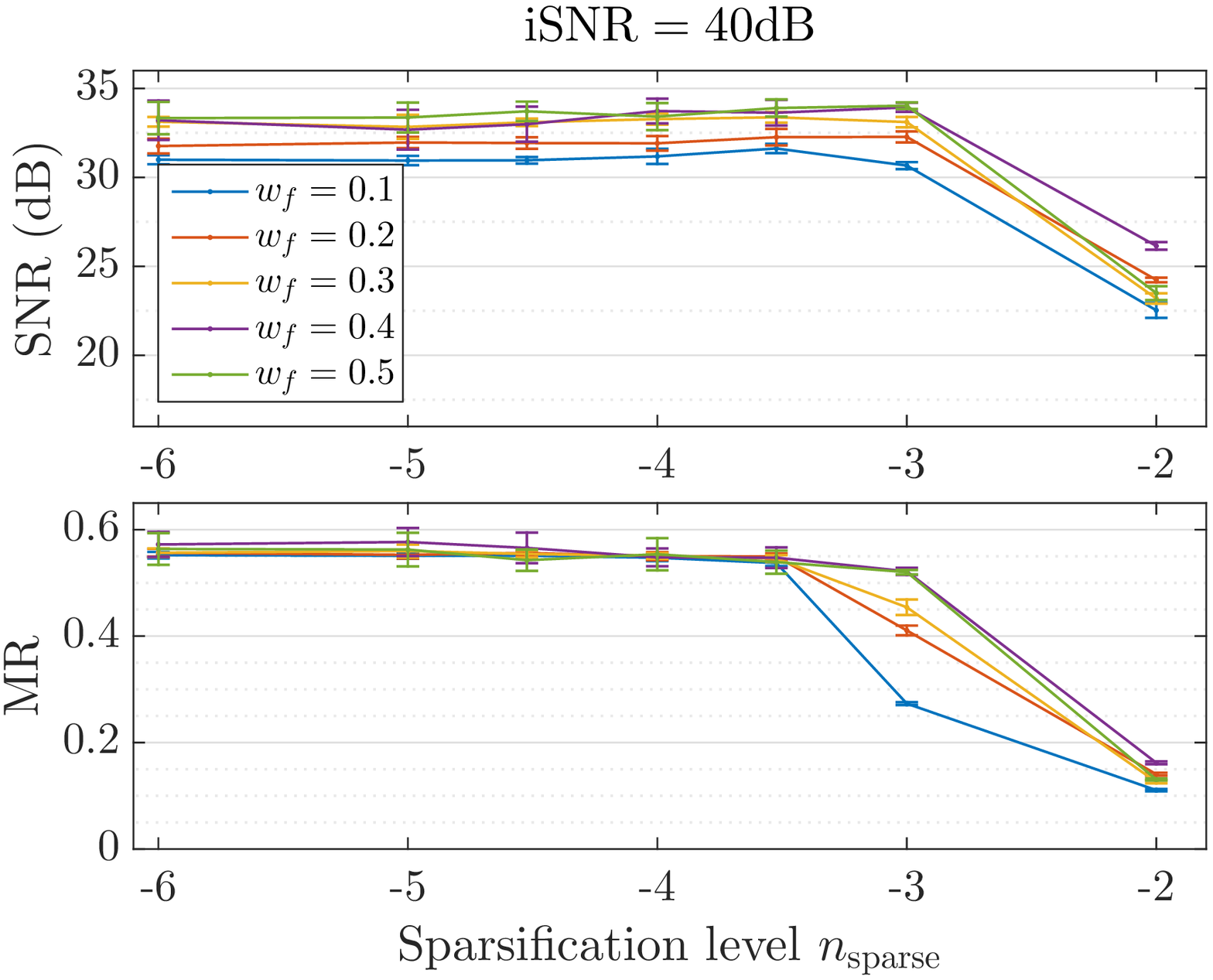}
 \caption{\small {Results of the operator $\hat{\bm C}$ sparsification tests for different sparsification levels $n_{\rm sparse}$ ($x$-axis) and iSNRs 30dB and 40dB. Top: SNR metric, bottom: MR metric ($y$-axis). Different colours indicate the fraction $w_f$ characterising the range of the $w$-rates. Data points correspond to the average over 10 noise realisations, error bars indicate their corresponding standard deviation.}}
 \label{fig_C_sparse_SNR40}
	\end{minipage}
	\begin{minipage}{0.99\linewidth}
	\vspace{3mm}
	\centering
 \includegraphics[width=0.75\textwidth]{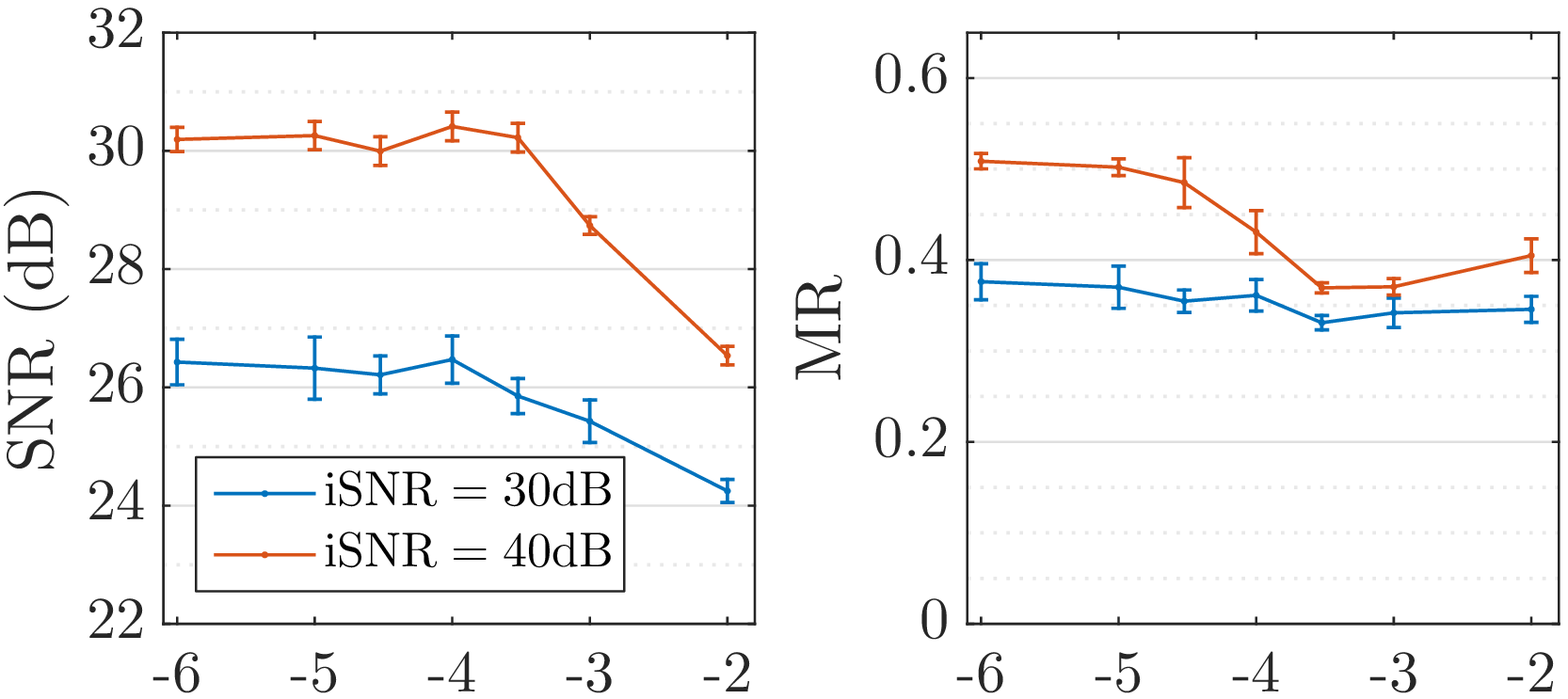}
 \caption{\small Results of the operator $\hat{\bm C}$ sparsification tests using the natural $w$-rates with iSNRs 30dB and 40dB. Left: SNR metric, right: MR metric ($y$-axis) as functions of the sparsification level $n_{\rm sparse}$ ($x$-axis). }
 \label{fig_C_worig}
 \vspace{3mm}
	\end{minipage}
\end{figure}

In Fig.~\ref{fig_Csparsity}, we show the sparsity ratio (SR) of the operator $\hat{\bm C}$ as a function of the sparsification levels $n_{\rm sparse}$. It is clear that increasing the sparsification levels results in a significant decrease of the operator's SR for all the considered ranges of the $w$-rates, even though the studied energy loss is, at most, two orders of magnitude smaller than the original energy. Naturally, the sparsity ratio of the $w$-projection $\tilde{\bm G}$ is of the same order as that of the Chirp operator $\hat{\bm C}$, given that the de-gridding kernels are of fixed small support $P =4\times 4$. } 

\begin{figure*}
\includegraphics[width=0.98\textwidth]{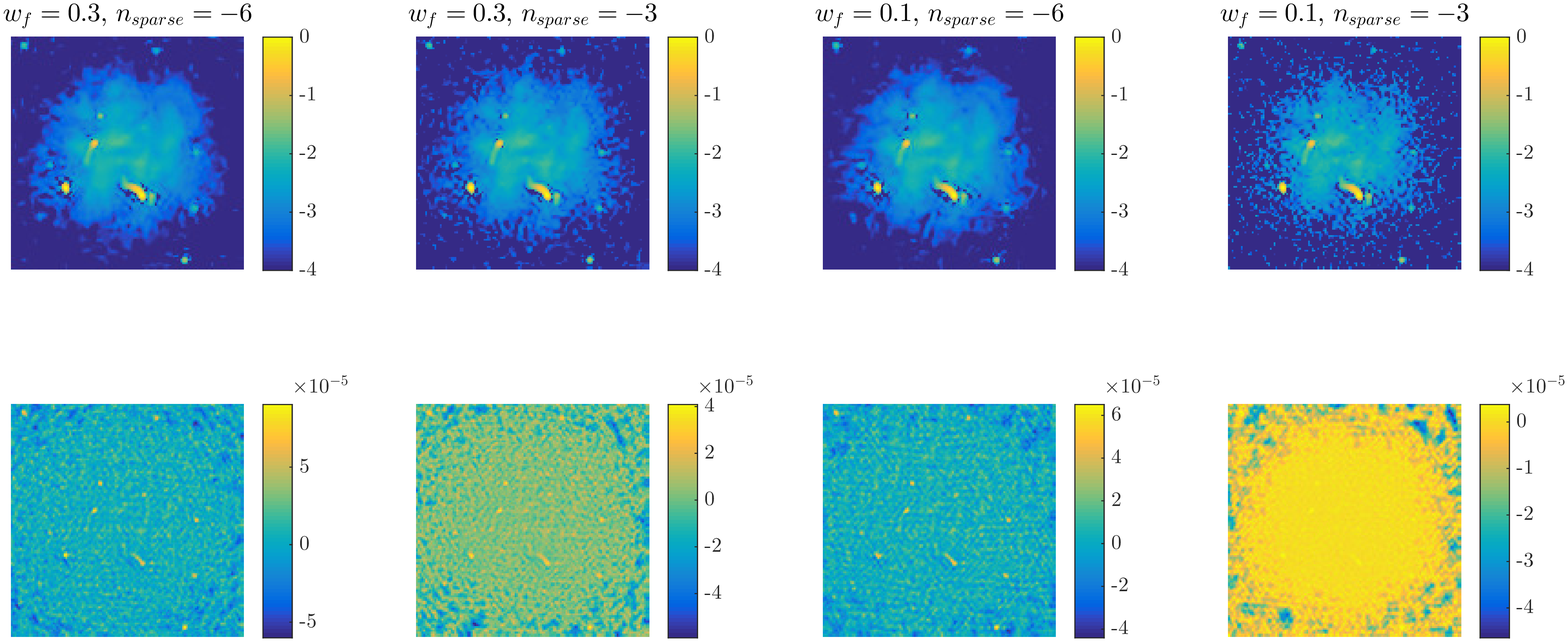}
\caption{\small Results of the sparsification of the Chirp operator $ \hat{\bm C}$  for $\rm iSNR=40dB$. Top: reconstructed images $\bar{\bs{x}}$, bottom: residual images $\bar{\bs{\Phi}}^{\dagger} (\bs{y}-\bar{\bs{\Phi}} \bs {\bar x})$ obtained by applying their corresponding sparse measurement operator $\bar{\bs{\Phi}}$. From left to right, reconstructed and residual images for $w_f=0.3$ ; $n_{\rm sparse}=-6$ (estimated image quality: $\rm SNR=32.54dB$, $\rm MR=0.55$, residual \textit{gaussianity}: $\rm skewness =-0.13$, $\rm kurtosis=5.08$ ), $w_f=0.3$ ; $n_{\rm sparse}=-3$ (estimated image quality: $\rm SNR=32.86dB$, $\rm MR=0.46$, residual \textit{gaussianity}: $\rm skewness=-1.09$, $\rm kurtosis=5.72$ ), $w_f=0.1$ ; $n_{\rm sparse}=-6$ (estimated image quality: $\rm SNR=30.58dB$, $\rm MR=0.54$, residual \textit{gaussianity}: $\rm skewness=-0.11$, $\rm kurtosis=4.67$) and $w_f=0.1$ ; $n_{\rm sparse}=-3$ (estimated image quality: $\rm SNR=30.53dB$, $\rm MR=0.27$, residual \textit{gaussianity}: $\rm skewness=-2.07$, $\rm kurtosis=7.98$).}
\label{fig_C_sparse_SNR40_images}
\end{figure*}

In Fig.~\ref{fig_C_sparse_SNR40}, we display the reconstruction quality of the estimated images $\bar{\bs x}$, in terms of SNR and MR as functions of the sparsification levels $n_{\rm sparse}$. Results are shown for the different $w_f$ considered, describing the different amplitudes of the $w$ values. For additive noise on the generated visibilities characterised by ${\rm iSNR}=30$dB, the SNR is not affected up to sparsification level smaller than $n_{\rm sparse}^* = -3$, independently of the $w$ range. Similar behaviour is observed by the MR metric. Note that, with respect to the sparsity of the $w$-projection operator, obtained with the least energy loss corresponding to $n_{\rm sparse} = -6$, the sparsity ratio up to the level $n_{\rm sparse} = -3$ has decreased by at least $50\%$ for $w_f = 0.5$ and $83\%$ for $w_f=0.1$. This constitutes a significant decrease in memory requirements, leading to a fast application of the measurement operators.
For sparsification level $n_{\rm sparse} =-2$ and energy loss on each row $ E^{\rm{lost}}_{\ell} = 10^{-2}$, both SNR and MR decrease significantly. This is due to the fact that model errors of radio interferometric measurements introduced by the sparsification of the operator $\hat{\bm C}$ hence the $w$-projection operator $\tilde{\bm G}$, are larger than the noise level on the measurements.
For $\rm iSNR=40dB$, the sparsification does not hamper the reconstruction quality up to sparsification level $n_{\rm sparse}^* = -3.5$. Note that, while this effect is not reflected in the SNR plot (Fig.~\ref{fig_C_sparse_SNR40}, upper panel), as the metric is also stable up to $n_{\rm sparse} = -3$, the MR metric (Fig.~\ref{fig_C_sparse_SNR40}, lower panel) is however stable up to $n_{\rm sparse}^*=-3.5$, for higher levels a significant decrease is observed (\textit{e.g.} $50\%$ decrease for $w_f=0.1$). 

We have also considered the sparsification of the Chirp operator $\hat{\bm C}$ for the natural $w$-rates of the  ASKAP $uv$-coverage. Most of the $w$ values are extremely small, resulting in flat chirps. Nevertheless, the reconstruction quality is affected for high sparsification levels (see Fig.~\ref{fig_C_worig}). In fact, for ${\rm iSNR}=30,40~\rm dB$ a decrease of the SNR by almost $\rm 2dB$ and $\rm 4dB$, respectively, is observed for the sparsification level $n_{\rm sparse}=-2$, where $83\%$ decrease of the SR of the $w$-projection operator $\tilde{\bm G}$ is reached. This is consistent with our finding at the smallest simulated $w$-rates (\textit{e.g.} $w_f=0.1$). The drop in the image quality is mainly exhibited as artefacts surrounding the faint diffuse central structure. 

Large errors on the model of radio interferometric measurements introduced by applying the sparsification of the operator $\hat{\bm C}$ and consequently the $w$-projection operator $\tilde{\bm G}$ result in estimated images with high artefacts. In fact, the convolution kernels of $\tilde{\bm G}$, being compact and hence not taking into account energy coming from distant Fourier modes, give rise to over-fitting of the probed Fourier modes. This effect can be shown through the inspection of Fig.~\ref{fig_C_sparse_SNR40_images}, where for the sparsification level $n_{\rm sparse} = -3$, the recovered images present strong artefacts consisting in large number of spurious point sources around the faint diffuse central structure. The effect is also reflected on the residual images when inspecting their histograms. More precisely, through the measures of skewness (this measures the lack of symmetry of the histogram) and kurtosis (this quantifies if the distribution is heavy-tailed or light-tailed relative to a normal distribution, in the latter case $\rm kurtosis = 3$), we found that the residual images with high sparsification level depart from a Gaussian noise. Moreover, for such high sparsification levels, we noticed that ADMM did not reach the bound $\epsilon$ on the $\ell_2$ ball that is set with respect to the noise statistics. Naturally, when model errors are larger than the noise level, solving  the constraint version of the minimisation problem (\ref{eq:min-pb}) results in data over-fitting due to the  inaccurate bound on the data fidelity term. 

In terms of computational time, the sparsification of the chirp kernels resulting in sparse measurement operators leads to significant acceleration of the image reconstruction step while preserving its quality. For instance, almost $45\%$ drop of the computational time is observed for tests with $w_f=0.5$ when considering a sparsification level $n_{\rm sparse}^*=-3.5$, \textit{i.e.,} a preserved energy $E^{\rm sparse}=0.9997$ with respect to the computational time for sparsification level $n_{\rm sparse}=-6$ with energy $E^{\rm sparse}=0.999999$. In general, for all the studied $w$ ranges, a decrease of at least $40\%$ of the computational time is observed for the same sparsification level $n_{\rm sparse}^*=-3.5$.
\subsection{Direct sparsification of the $w$-projection operator}
Adopting similar settings to the previous section, we study the sparsification technique to each row $\tilde{\bs{g}}_\ell$ of the operator ${\tilde{\bm G}}$ preserving the energy $E^{\rm sparse}_{\ell} = (1-{10}^{n_{\rm sparse}})\times E^{\rm{total}}_{\ell} $ with $n_{\rm sparse} \in \{-6, -5, -4, -3.5, -3, -2.5,-2\}$ denoting the sparsification level. We consider various $w$-rates within 5 different ranges, such that $ \forall ~{1\leq \ell \leq M},~|w_{\ell}|\leqslant w_f \times w^*$, with $w_f\in\{0.1, 0.2, 0.3,0.4, 0.5\}$. We also consider the original, un-scaled $w$ values given by the baseline distribution.
\begin{figure}
	\centering
	\begin{minipage}{0.99\linewidth}
	\vspace{3mm}
	\centering
	\includegraphics[width=0.75\textwidth, height=0.4\textwidth]{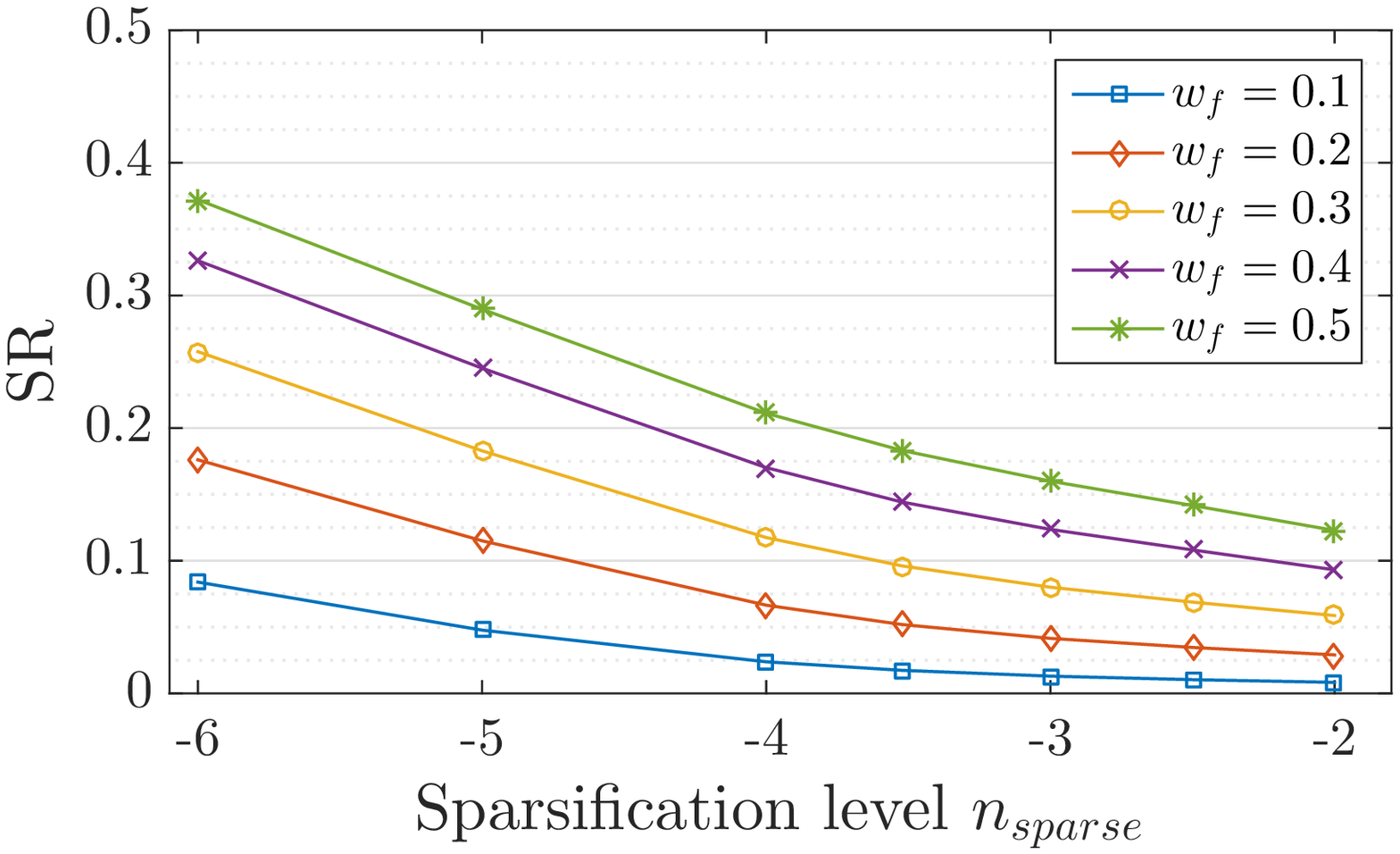}
\caption{\small Sparsity ratio ($y$-axis) of the $w$-projection operator $\tilde{\bm G}$, to which sparsification is directly applied using different sparsification levels $n_{\rm sparse}$ ($x$-axis). }
\label{fig_Gsparsity}
\vspace{3mm}
 \end{minipage}
	
	\begin{minipage}{0.99\linewidth}
	\centering
 \includegraphics[width=0.85\textwidth]{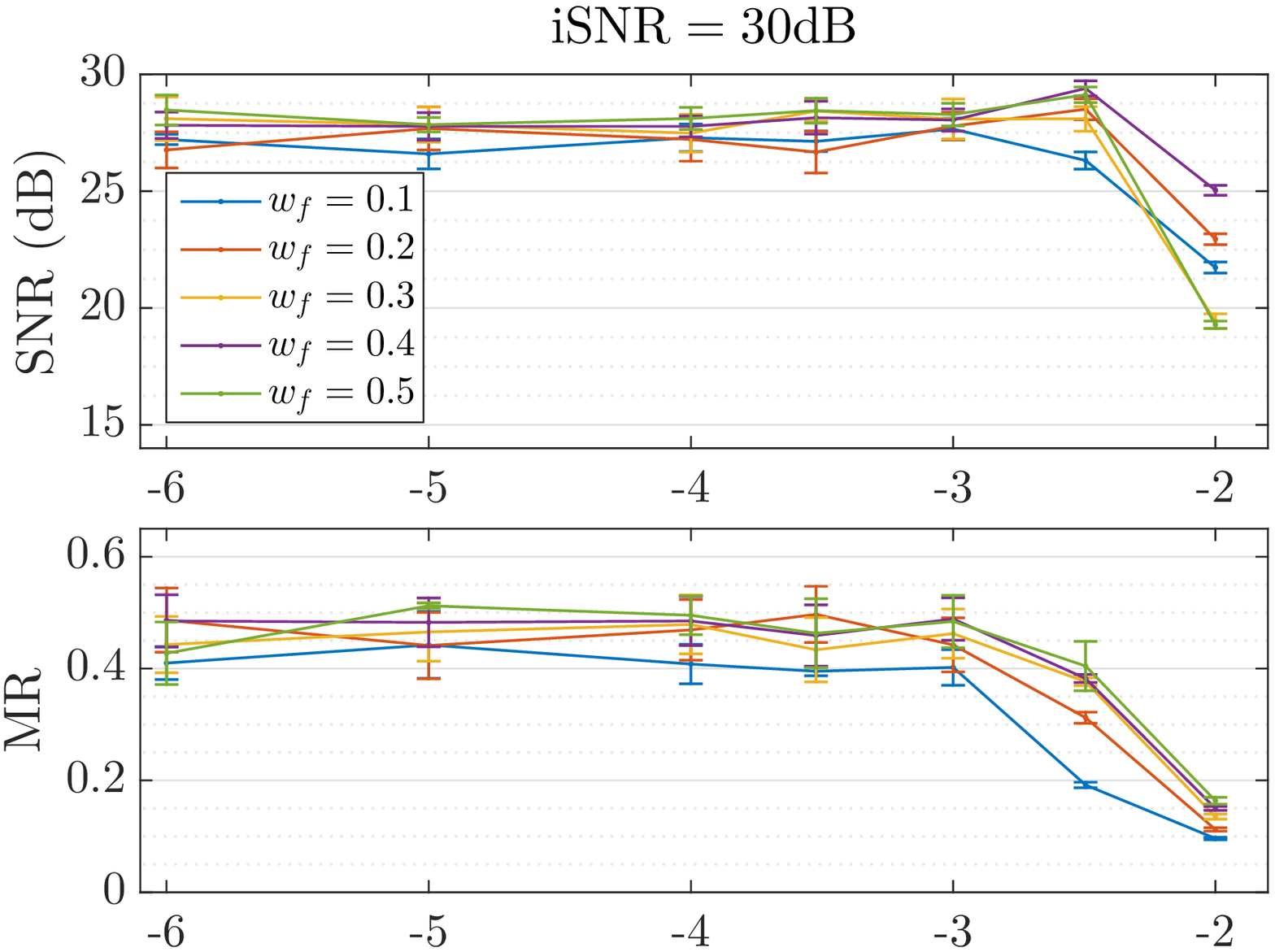}
	\end{minipage}
	~\\
	~\\		
	\begin{minipage}{1\linewidth}
	\centering
 \includegraphics[width=0.85\textwidth]{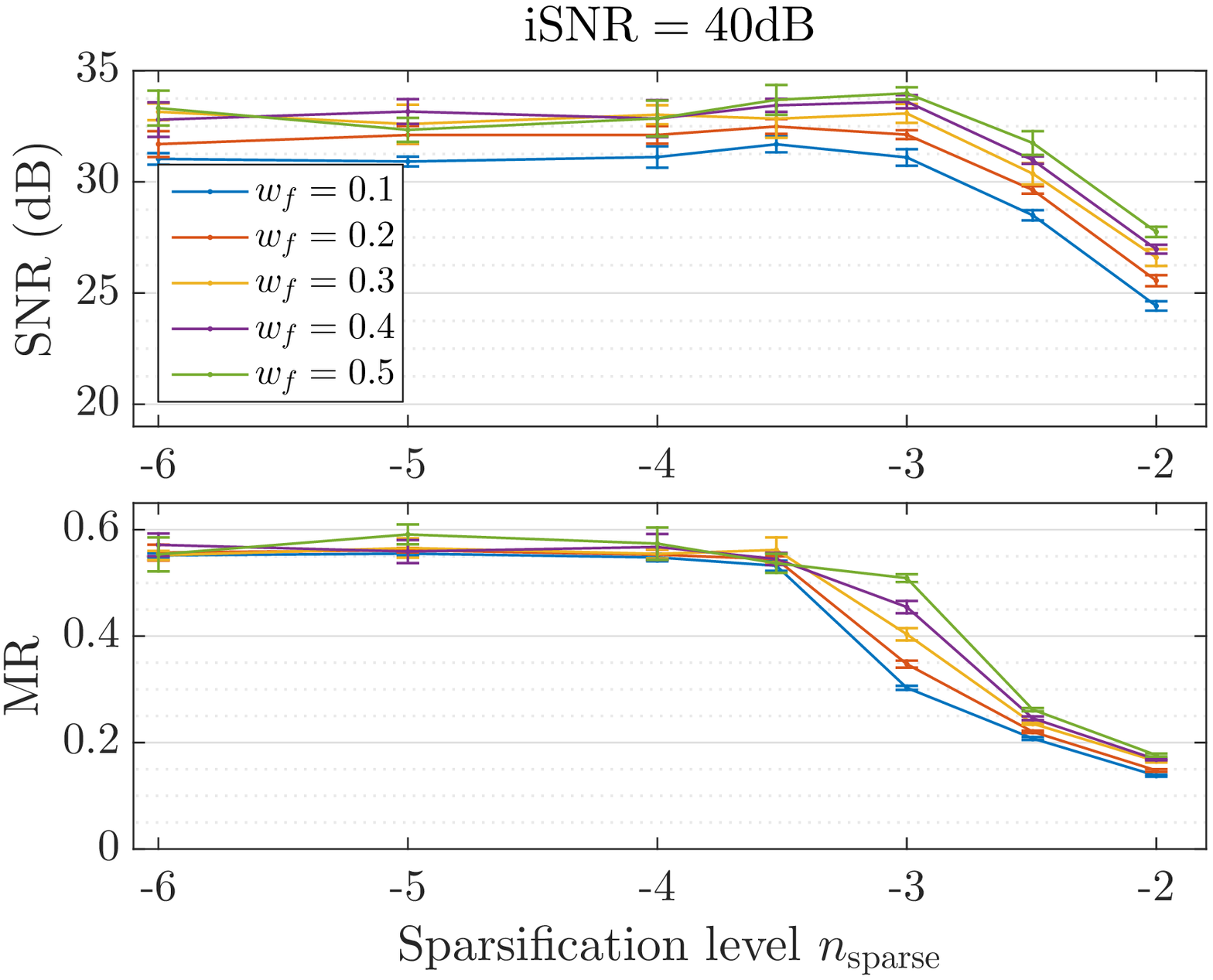}
\caption{\small {Results of the operator $\tilde{\bm G}$ sparsification tests for different sparsification levels $n_{\rm sparse}$ ($x$-axis) and {iSNRs 30dB and 40dB}. Top: SNR metric, bottom: MR metric ($y$-axis). Different colours indicate the fraction $w_f$ characterising the range of the $w$-rates. Data points correspond to the average over 10 noise realisations, error bars indicate their corresponding standard deviation.}}
\label{fig_G_sparse_SNR40}
	\end{minipage}
	\begin{minipage}{0.99\linewidth}
	\vspace{3mm}
	\centering
  \includegraphics[width=0.75\textwidth]{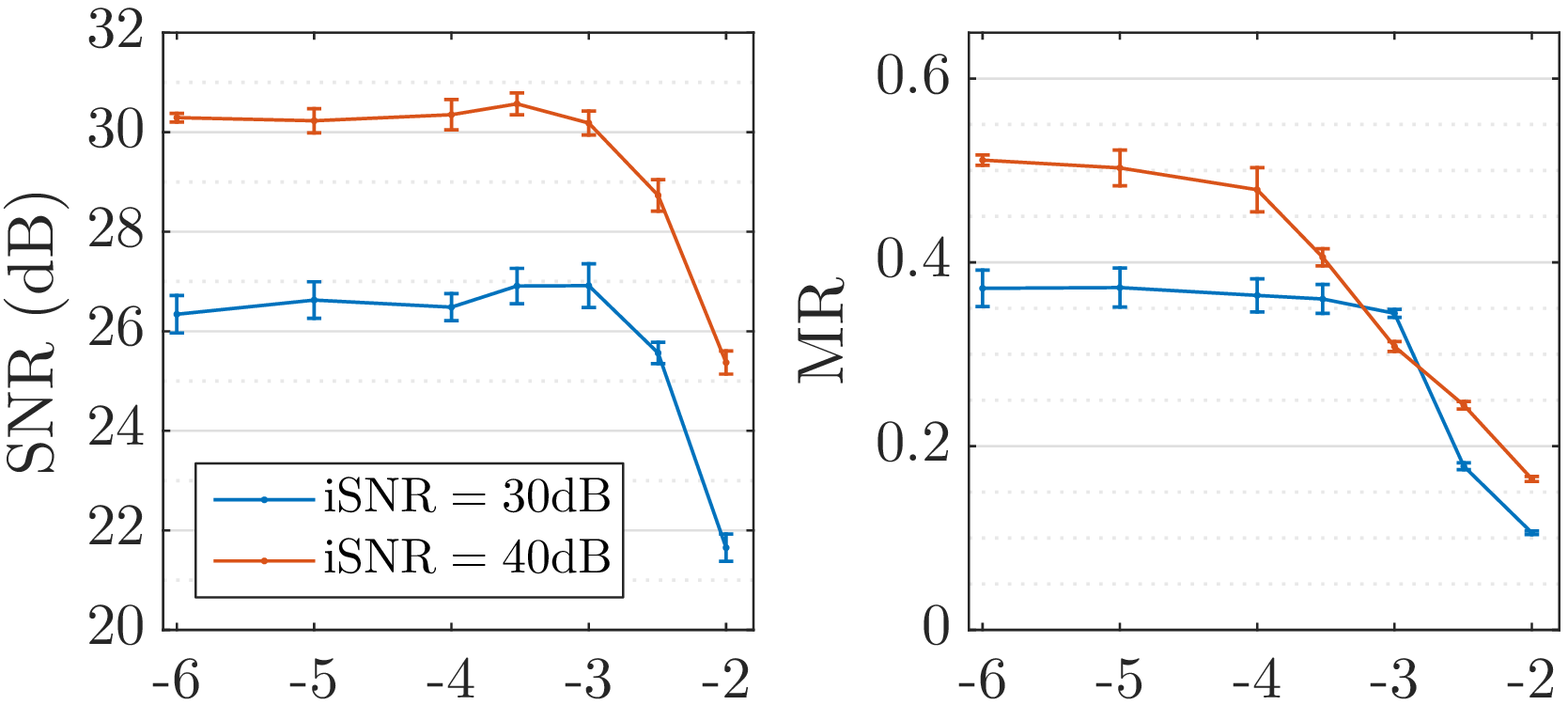}
 \caption{\small {Results of the operator $\tilde{\bm G}$ sparsification tests using the natural $w$-rates with iSNRs 30dB and 40dB. Left: SNR metric, right: MR metric ($y$-axis) as functions of the sparsification level $n_{\rm sparse}$ ($x$-axis). }}
  \label{fig_G_worig}
  \vspace{3mm}
	\end{minipage}
\end{figure}
In Fig.~\ref{fig_Gsparsity}, we show the sparsity ratio of the operator $\tilde{\bm G}$, where a drop is clearly visible for increasing sparsification levels. For instance, considering $w_f =0.5$, the sparsity ratio for $n_{\rm sparse}=-6$ is 0.38 and decreases by $57\%$ for $n_{\rm sparse} =-3$. Also in the case $w_f =0.1$, the sparsity ratio for $n_{\rm sparse}=-6$ is 0.08, and decreases by $87\%$ for $n_{\rm sparse} =-3$. 
In Fig.~\ref{fig_G_sparse_SNR40}, first and second panels, the results are obtained with ${\rm iSNR}=30$dB. For the $w$ ranges considered, the SNR values are unaffected by the sparsification of the operator $\tilde{\bm G}$ up to $n_{\rm sparse} =-2.5$. The MR is however stable up to $n_{\rm sparse}^*=-3$. Note that, the values of the SNR at level $n_{\rm sparse}\geq -2.5$ are biased by the reconstruction of the brightest sources in the GC image. In this case, the MR metric allows for a better judgement of the reconstruction quality. For high sparsification level $n_{\rm sparse}=-2$, a significant drop of both SNR and MR is observed. It is worth mentioning that the high SNR values obtained with $n_{\rm sparse}=-2.5$ with respect to smaller levels are promoted by the over-fitting effect due to sparsification. In fact, this effect allows for a better recovery of the higher spacial-frequency content in the image coming from the brightest compact sources. Yet, this comes at the expense of the recovery of the faint extended emission, which is mainly contributing to the low spacial-frequency content. Results of the sparsification with noise level on the visibilities of $\rm {iSNR}=40$dB are shown in Fig.~\ref{fig_G_sparse_SNR40}, third and fourth panels. Similar behaviour is obtained and the quality of the image reconstruction is preserved up to the sparsification level $n_{\rm sparse}^*=-3.5$.

We apply this sparsification strategy for the natural  $w$-rates. In this test case, the sparsification of the operator $\tilde{\bm G}$ can also act on the de-gridding kernels; for extremely small $w$ values where the chirp kernel is reduced to a Dirac. Note that sparsifying the Chirp operator $\hat{\bm C}$ as described in the previous section preserves the de-gridding kernels as it is performed before the convolution of the two kernels. Interestingly, when it is the case, we notice that the reconstruction quality is not prone to degradation up to sparsity levels $n_{\rm sparse}^*= -4$ for ${\rm iSNR}=40$dB and $n_{\rm sparse}^*= -3$ for $\rm iSNR=30$dB while decreasing the SR of the operator $\tilde{\bm G}$ by $37\%$ and $48\%$ respectively (see Fig.~\ref{fig_G_worig}). For higher sparsification levels a significant drop in the reconstruction quality is observed, that is translated in the image as large artefacts hampering the recovery of the faint structures, and in particular the non-recovery of the faint and extended central emission.  

Regarding the computational time, it is generally significantly reduced with respect to the tests corresponding to the minimal sparsification level $n_{\rm sparse}=-6$. A decrease by around $30\%$ is noticed for $w_f=0.1$ and at least $40\%$ for $w_f>0.1$ when adopting the sparsification level $n_{\rm sparse}^*=-3.5$. No noticeable increase for the test case adopting the original $w$-rates. 

When compared with the previous sparsification strategy applied on the Chirp operator, the direct sparsification of the $w$-projection operator leads to lower SR of its sparse approximation obtained by the highest and quality preserving sparsification levels; $n_{\rm sparse}^* =-3$ for iSNR$=30$dB and $n_{\rm sparse}^* =-3.5$ for iSNR$=40$dB. We also notice a decrease in the computational time for these levels. This suggests that the direct sparsification strategy not only lowers memory requirements but also the computational time as well. In particular, direct sparsification of the $w$-projection operator $\tilde{\bm G}$ with energy loss fraction $\gamma=10^{-4.5}$ has shown to preserve image reconstruction quality for ${\rm iSNR}$ up to 40dB on simulated visibilities. Therefore, we consider this case as our fiducial setting for a sparse measurement operator in the next section.
\subsection{Revisiting the spread spectrum effect on image reconstruction quality}
\begin{figure}
\centering
\includegraphics[scale=0.24]{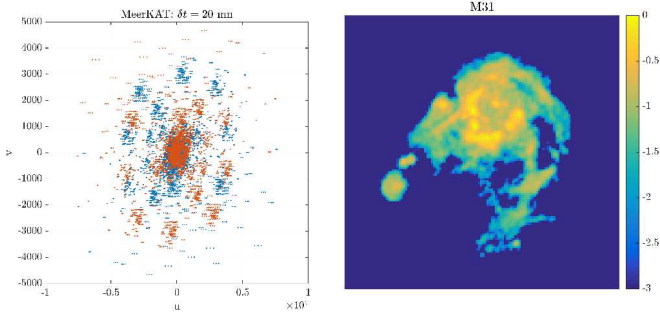}
\caption{{\small{Left: adopted MeerKAT $uv$-coverage, having $M =6048$ measurements coloured in red; the other half of the coverage, coloured in blue, is obtained with hermitian symmetry. Right: the galaxy image $\rm{M31}$ of size $128\times128$ pixels, displayed in $\rm{log}10$ scale. }}}
\label{fig_emeerkat}
\end{figure}

\begin{figure}
\centering
\includegraphics[width=0.23\textwidth]{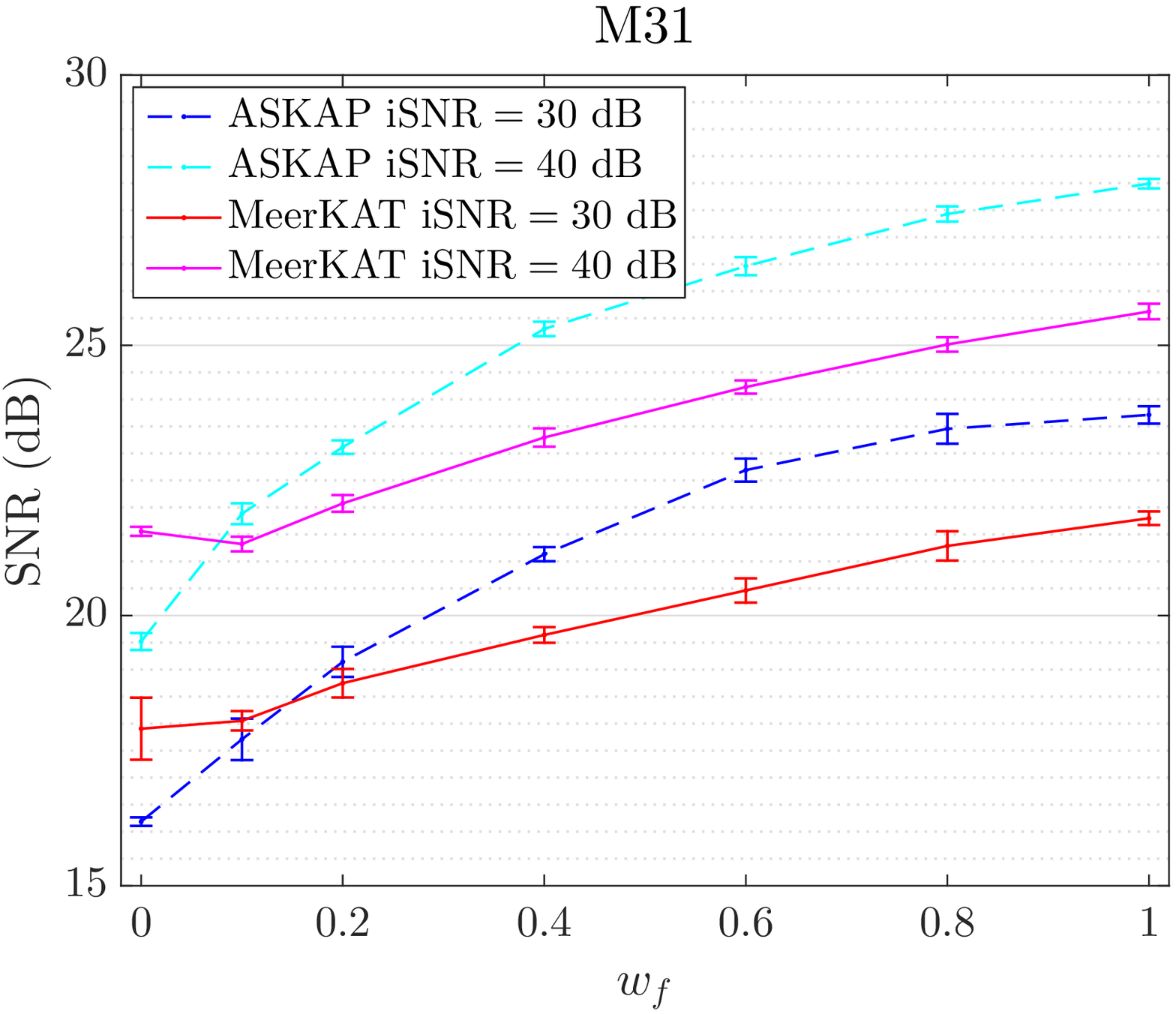}
\includegraphics[width=0.23\textwidth]{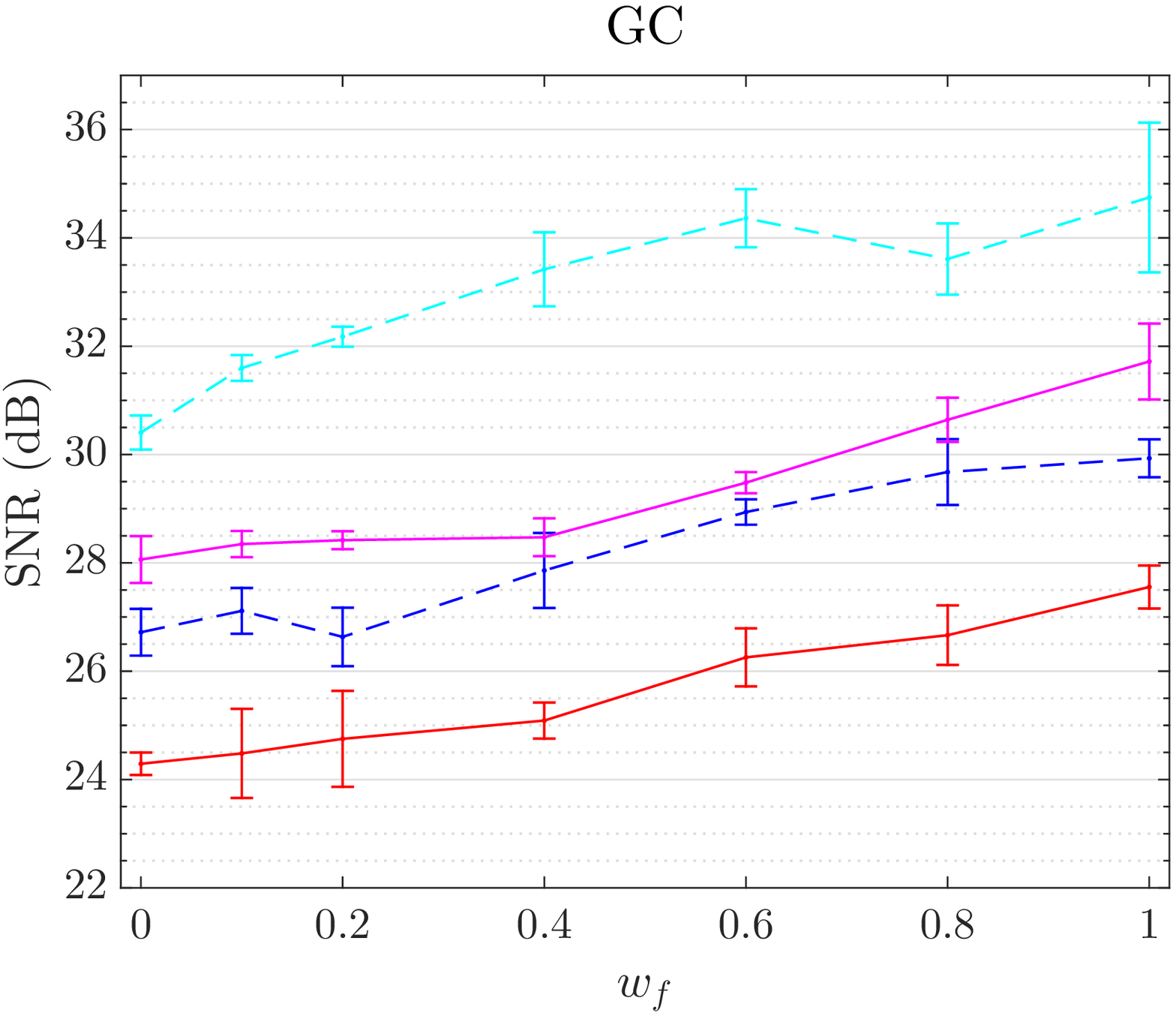}
\caption{\small SNR of the reconstructed images as a function of $w_f$ for the {$\rm{iSNRs}=30, 40{\rm dB}$} and two different $uv$-coverages of ASKAP and MeerKAT. Left: M31 image, right: GC galaxy cluster image. Data points correspond to the average over 10 noise realisations, and the error bars are derived from the standard deviation of this set of runs.}
\label{Final_M31_GC}
\end{figure}

We highlight the general quality enhancement of the reconstructed images by adopting the $w$-projection algorithm in the framework of convex optimisation for increased $w$-rates. To demonstrate the versatility of our study, we have extended our simulation settings by further adopting the image of the galaxy M31 of size $128\times128$ pixels. We have also considered a simulated $uv$-coverage from the antenna configuration of the Karoo Array Telescope (MeerKAT)\footnote{http://www.ska.ac.za/gallery/meerkat/}, a precursor of the SKA. The coverage is obtained, 1 hour observation time and time interval $\delta{\rm t}=20$ minutes, pointing at declination $0d0m0s$ and right-ascension $10d0m0s$. The total number of $\bs u$ points is $M=6048$. Both, the image of M31 and the MeerKAT coverage are displayed in Fig.~\ref{fig_emeerkat}. We apply direct sparsification of the $w$-projection operator with a sparsification level $n_{\rm sparse}=-3.5$ in the backward model. Results are displayed in  Fig.~\ref{Final_M31_GC}, where the evolution of SNR of the reconstructed M31 (left panel) and GC (right panel) images are shown as a function of the $w$-rates characterised by $w_f$. The output SNR improves with the increasing ranges of $w$-rates for all settings. That is in agreement with the previous studies of the spread spectrum effect. 
\section{Super-resolution promoted by the $w$-term}
\label{sec:SR}
In absence of $w$-modulation, the new imaging approaches based on compressive sensing and sparse representations, have shown remarkable improvement of the angular resolution of the reconstructed images when compared to the output restored images of CLEAN, these are limited by the instrumental PSF. Thanks to the sparsity-based prior and positivity imposed on the image to be recovered, this class of approaches has shown the ability to go beyond the instrument resolution limit. In presence of $w$-modulation, the resulting spread spectrum effect \citep{Wiaux09,Wolz2013} allows for probing radio interferometric measurements containing not only information on the sensed Fourier modes of the imaged sky but also information coming from the surrounding Fourier modes induced by the convolution effect of the $w$-modulation. In other words, this means that a radio interferometric measurement at a probed spacial frequency $\bs u$ contains energy that consists, not only of the Fourier component of the imaged sky at that specific spacial frequency $\bs u$, but also of a linear combination of the Fourier components of the imaged sky from the neighbouring spacial frequencies. This suggests that high frequency content of the signal, that is beyond the band limit of the interferometer, can be probed. Intuitively, this might yield to the recovery of super-resolved images, with bandwidth $B'_{\rm max}$ larger than the bandwidth of the array, that is given by the maximum projected baseline (see subsection \ref{ss:mo}). 

In this section, we study the potential of the $w$-term to allow for super-resolution on two test cases; the first consists of the study of super-resolution on the galaxy cluster image and the second the study of super-resolution allowing for the detection of separated point sources below confusion noise. The latter is a direct effect of the band limit of the radio interferometer. We re-emphasize that, the spread spectrum effect is not exclusive to the $w$-modulations; in fact it is fundamentally induced by DDEs as they are modelled through convolutional kernels in the Fourier domain, similarly to the $w$-term. Yet, unlike the $w$-terms, they are unknown. Therefore, they need to be estimated along with the unknown radio image, resulting in non-convex blind deconvolution problem. 
\subsection{Super-resolution on the Galaxy Cluster image}
We simulate observations using the GC image of size $N=128\times128$ as a ground-truth image and a realistic simulated $uv$-coverage of ASKAP antenna configuration for a total observation time of 1 hour, pointing at declination $-10d0m0s$ and right-ascension $0d0m0s$ with time spacing $\delta{\rm t}=10$ minutes, leading to $M=3780$ visibilities. Obtained visibilities are contaminated with additive white Gaussian noise with $\rm{iSNR}=40$dB following (\ref{eq:ra-problem}). Note that in this study, we consider simulations of the ground-truth image sensed with a radio interferometric configuration characterised with a band limit smaller than its full bandwidth $B$. Therefore, we generate $w$-modulations so that $B'_{\rm max}=\displaystyle\max_{{1\leq \ell \leq M}}{B_{w_{\ell}}}+{\parallel {\bs u_{\ell}}\parallel_2}$ does not exceed $B$, hence no up-sampling operator is required. 
\begin{figure*}
	\centering
 \includegraphics[width=1\textwidth]{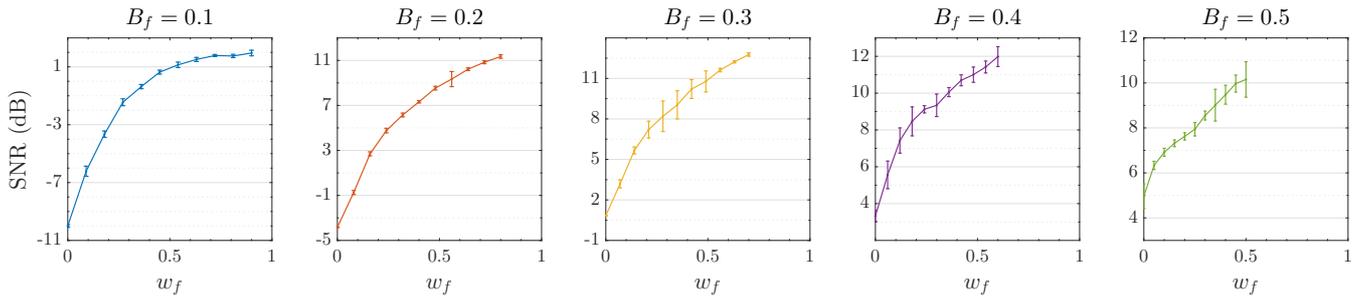}
\caption{\small SNR results of the super-resolution tests ($y$-axis) on the GC image in the Fourier domain non-probed by the sampling as a function of the $w_f$ ($x$-axis) that is characterizing the range of the generated $w$ values. Different colours indicate the scaling factor $B_f$ of the adopted $uv$-coverage. Data points correspond to the average over 10 different realisations of the $w$-rates and the noise, and the error bars are derived from its standard deviation.}
\label{fig_GC_SR_SNR40_F}
\end{figure*}
We perform a family of tests by down-scaling the simulated $uv$-coverage with factors $B_f \in\{ 0.1, 0.2 ,0.3, 0.4,0.5\}$. For every test, we set the probed bandwidth $\tilde{B}= B_f \times B$, where $B$ is the image full bandwidth. We also generate random $w$ values within ranges characterized by $w_f\in\{0.1,0.2,0.3,0.4,0.5,0.6,0.7,0.8,0.9,1\} \times w_f^*\times w^*$, where $w^*$ is the maximum $w$-rate inducing a chirp modulation with a bandwidth $B_c =B$. The value $w_f^*$ is the $w$-fraction allowing to probe the non-sensed band of the Fourier plane by the down-scaled $uv$-coverage with a factor $B_f$. In this case, $w_f^*$ induces a chirp modulation with a bandwidth $B_c= (1-B_f) \times B$.
 
Let $\bs z$ denote the Fourier components of the sky $\bs x$ at the non-sensed high frequency modes such that ${\bs z} =[\Re({{\bm M }{\bm F }{\bs x}});\Im({{\bm M }{\bm F }{\bs x}})]$. $\bm M\in \mathbb{R}^{N\times N}$ is a diagonal matrix with binary entries selecting the Fourier modes that are part of the non-probed band and $\bs z \in \mathbb{R}^{2N\times N}$, is the concatenation of the real and imaginary parts of the selected Fourier components ${\bm M }{\bm F }\bs x$. To quantify the super-resolution effect promoted by the $w$-modulation, we use the SNR metric on $\bs z$ as follows:
\begin{align*}
{\rm SNR} = 20 \log_{10} \frac{|| {\bs z}||_2}{|| {\bs z}-{\bar{\bs z} }||_2},
\end{align*}
where $\bar{\bs z}=[\Re({{\bm M }{\bm F }\bar{\bs x}});\Im({{\bm M }{\bm F }\bar{\bs x}})]$ and $\bar{\bs x}$ is the reconstructed image of the sky. 

\begin{figure}
 \includegraphics[width=0.4\textwidth]{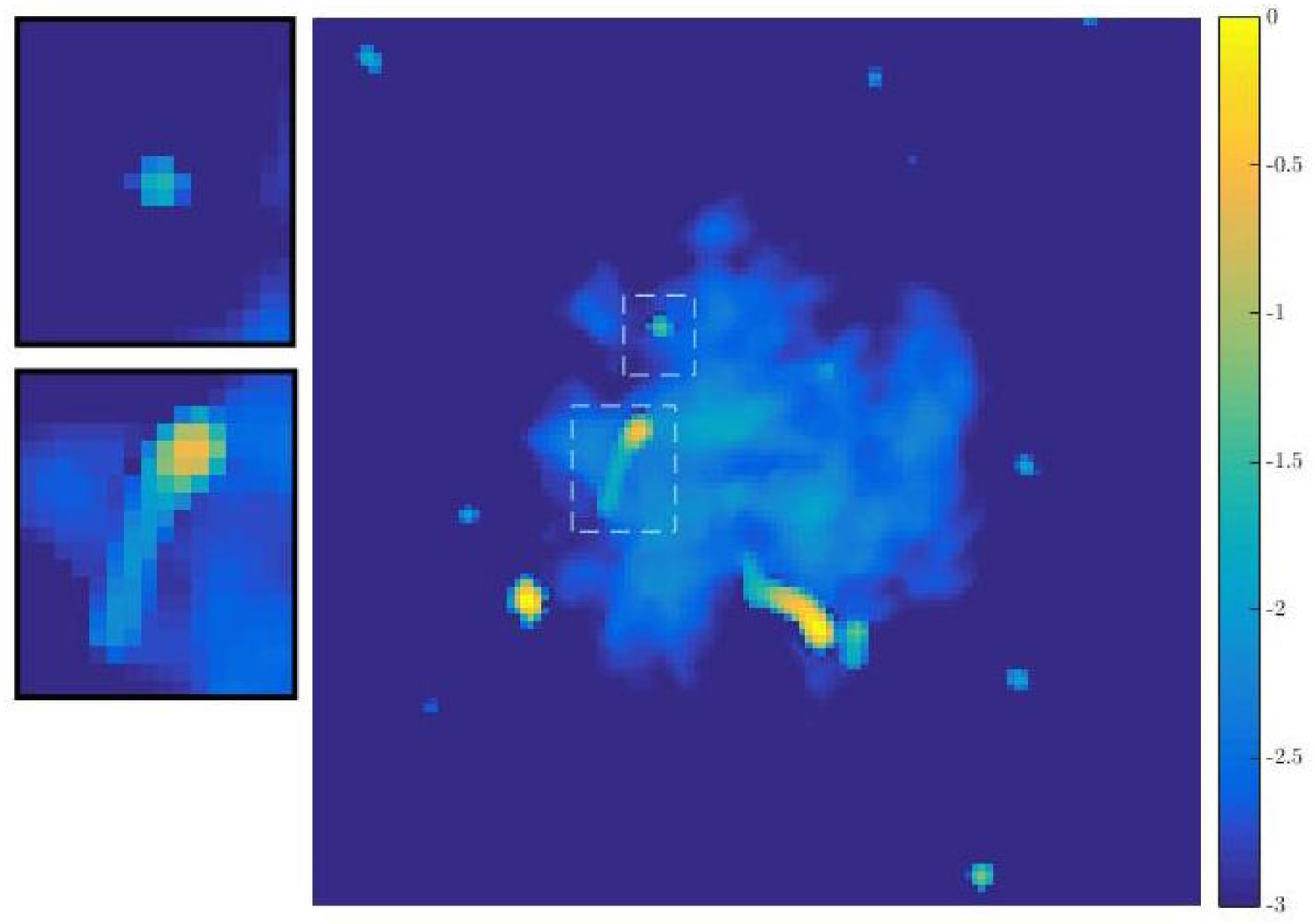}
 \includegraphics[width=0.4\textwidth]{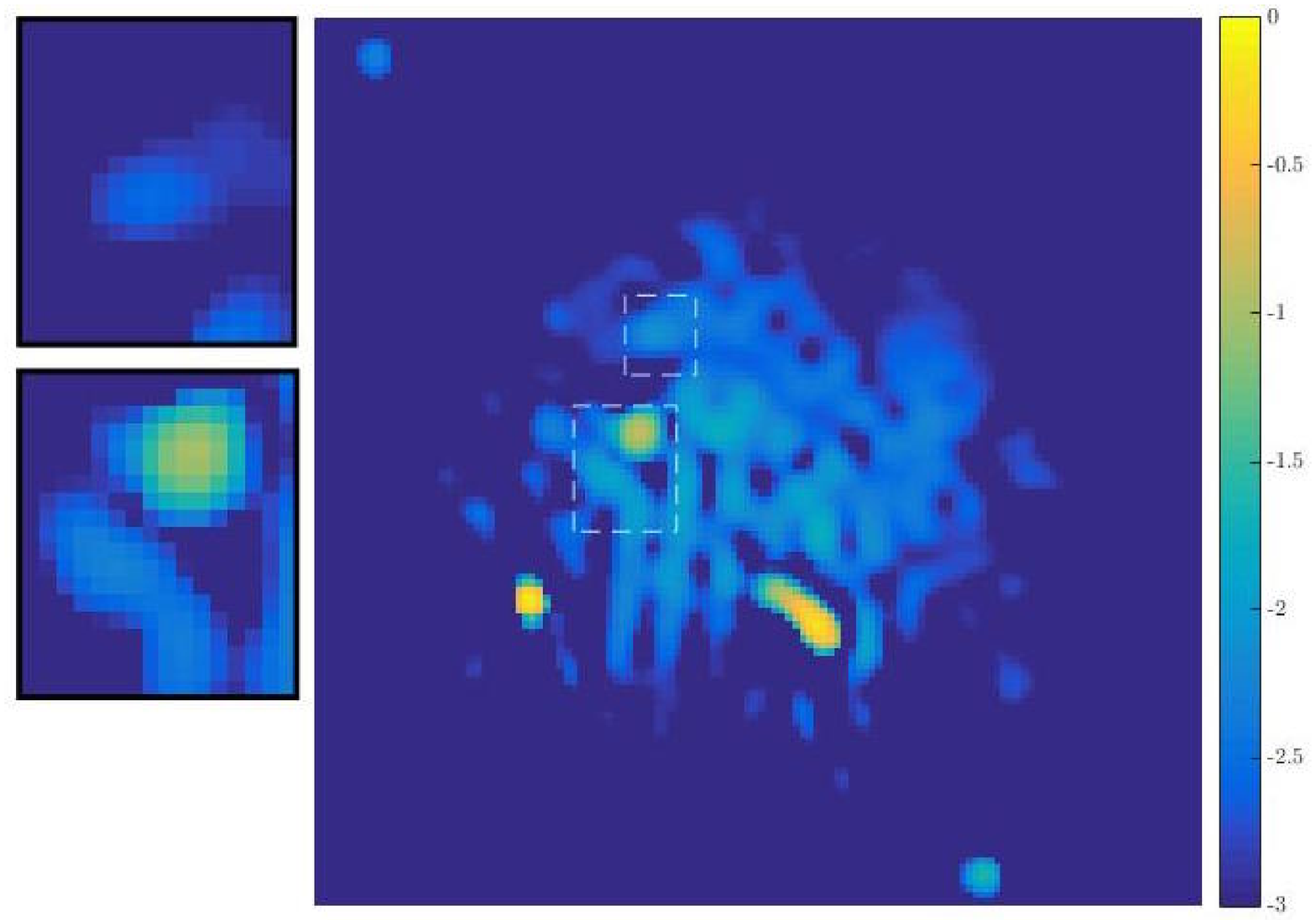}
 \includegraphics[width=0.4\textwidth]{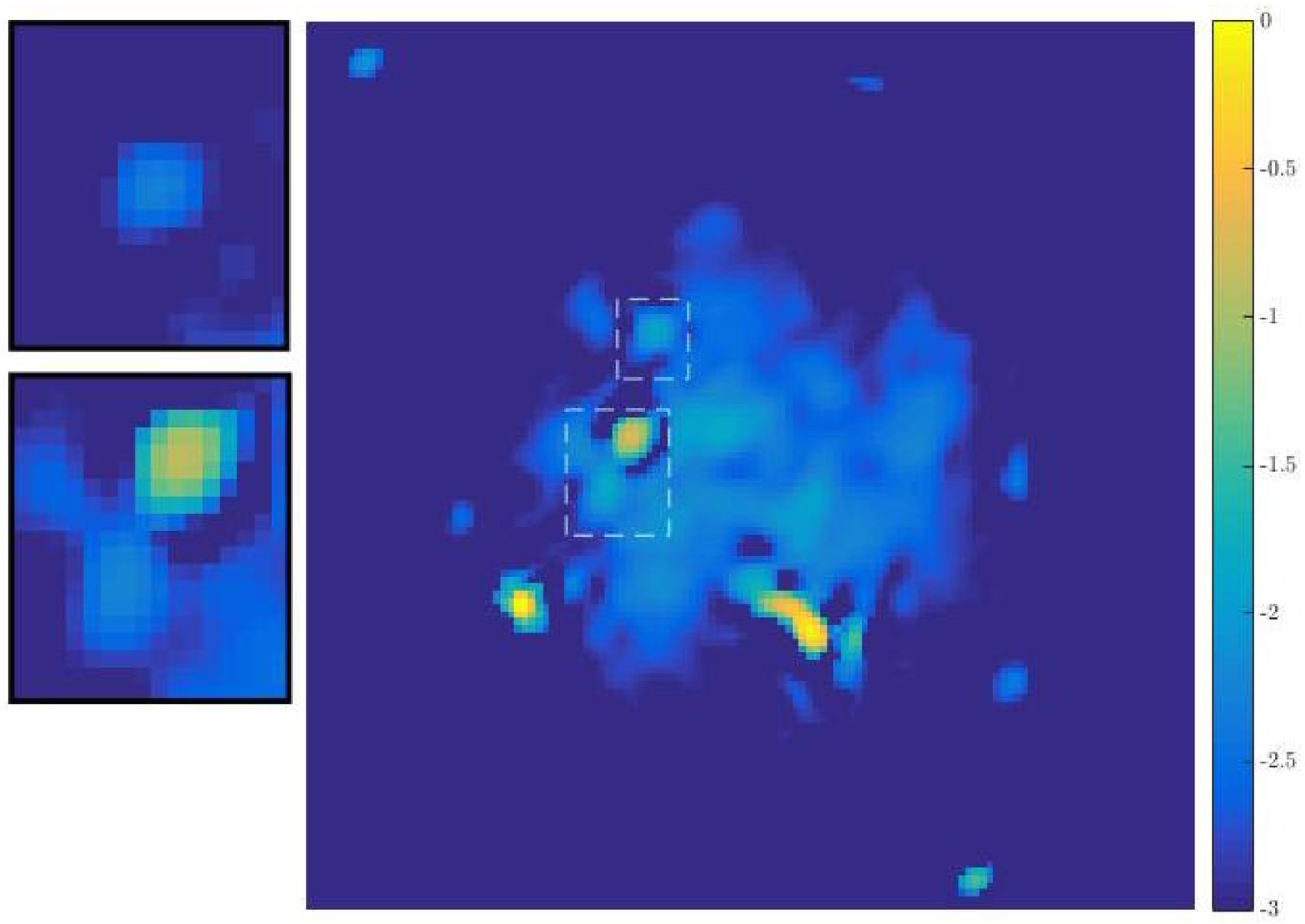}
 \includegraphics[width=0.4\textwidth]{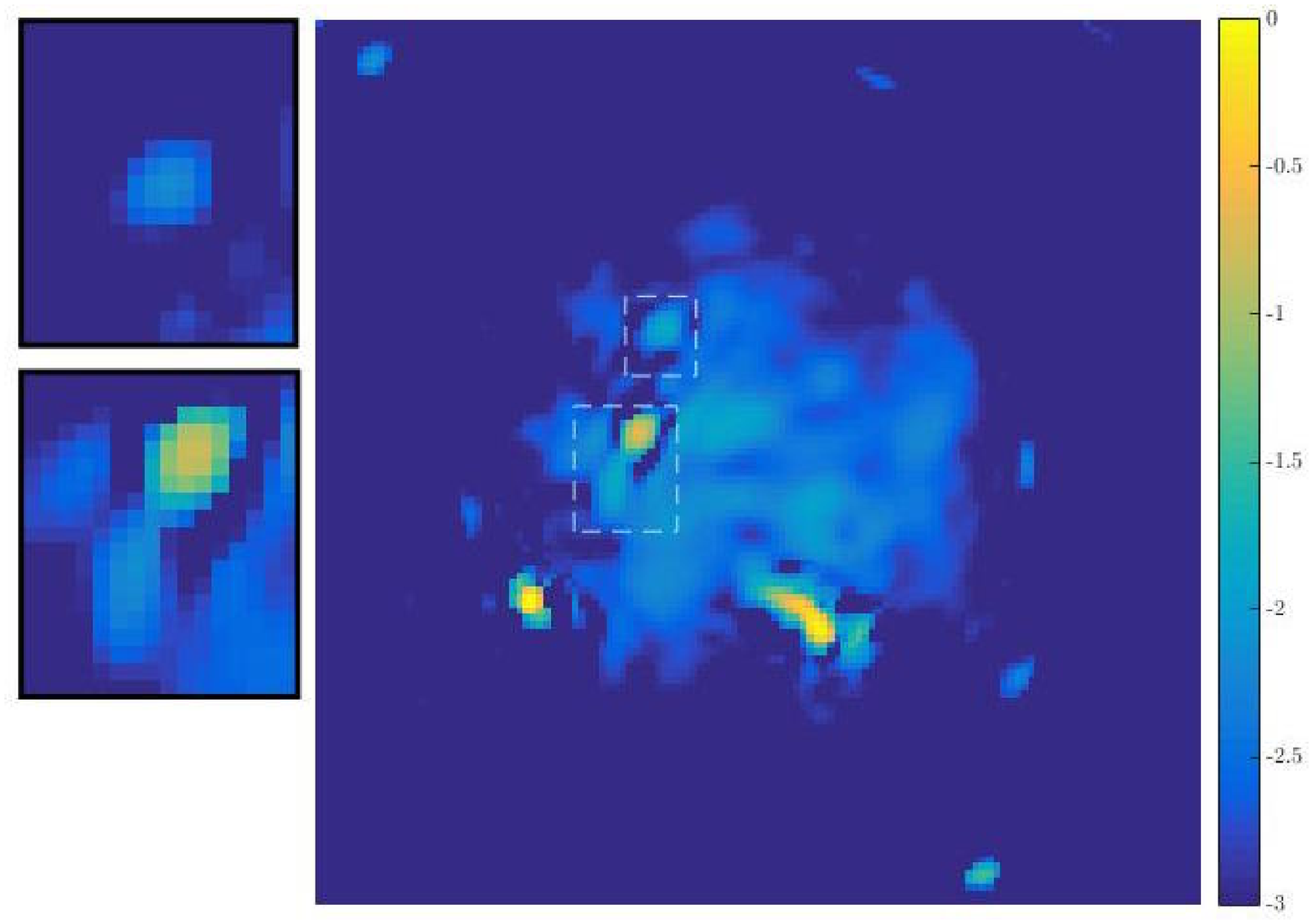}
\caption{\small Reconstructed images of the adopted $uv$-coverage with downscaling factor $B_f=0.3$, displayed in log10 scale. From top to bottom, ground-truth image and estimated images for $w$ ranges characterised respectively with $w_f=0,0.5,1$. Left: zoom on two sources of the images (note that different scale range is adopted for a better visualisation). When inspecting the bright galaxy in the GC image presenting a faint tail, it is clear that in the absence of $w$-modulation the tail is unresolved within the halo structure, while incorporating the $w$-modulation with large rates, a very good recovery of the tail is achieved. Same behaviour for the point source above it, which in absence of $w$-modulation is completely unresolved.}
\label{fig_im10}
	
\end{figure}

In Fig.~\ref{fig_GC_SR_SNR40_F}, we display the evolution of the SNR for the Fourier components  $\bs z$  belonging to the non-sensed band in the Fourier domain, as a function of the range $w_f$ characterising the $w$-rates. Clearly, the increase of the values of $w$ results in the improvement of the SNR for the various sensed bandwidths characterised by $B_f\in \{0.1,0.2,0.3,0.4,0.5\}$; more than 5dB increase of the SNR is observed with the highest $w_f= w_f^*\times w^*$, that is allowing to probe the full bandwidth of the image, when compared to no $w$-modulation ($w_f=0$). Moreover, the visual inspection of Fig.~\ref{fig_im10} confirms the super-resolution effect promoted by large $w$-modulations, in particular when inspecting compact sources present in the GC image.  In general, the SNR improvement is more noticeable for smaller sensed bandwidths. Note that, for $B_f \geq 0.4$, the frequency content of the image under scrutiny at the non-sensed Fourier modes is very small. Interestingly, in absence of the $w$-modulation, the positivity and sparsity priors allow for a small super-resolution effect, that is reflected in the obtained SNRs; 3.3dB and $4.96$dB for $B_f = 0.4$ and $B_f = 0.5$, respectively. 

\subsection{Point-source confusion limit}
Radio measurements are prone to point source confusion due to the band limit of the radio interferometer. This can be reflected in the so-called confusion noise, that is a noise-like background consisting of large number of very faint and slightly above the noise level point-like sources and which are below the resolution limit, thus unresolved. The confusion can also occur in the case of bright unresolved point sources lying on a diffuse background, such case can represent a supernova remnant or protostars in a nebula environment. In this section, we study the ability of the $w$-term to promote super-resolution in particular the ability to separate two unresolved point sources, hence decreasing the confusion limit due to the band limit of the radio interferometer.

\begin{figure*}
\begin{minipage}{1\linewidth}

	\centering
 \includegraphics[width=1\textwidth, clip=true]{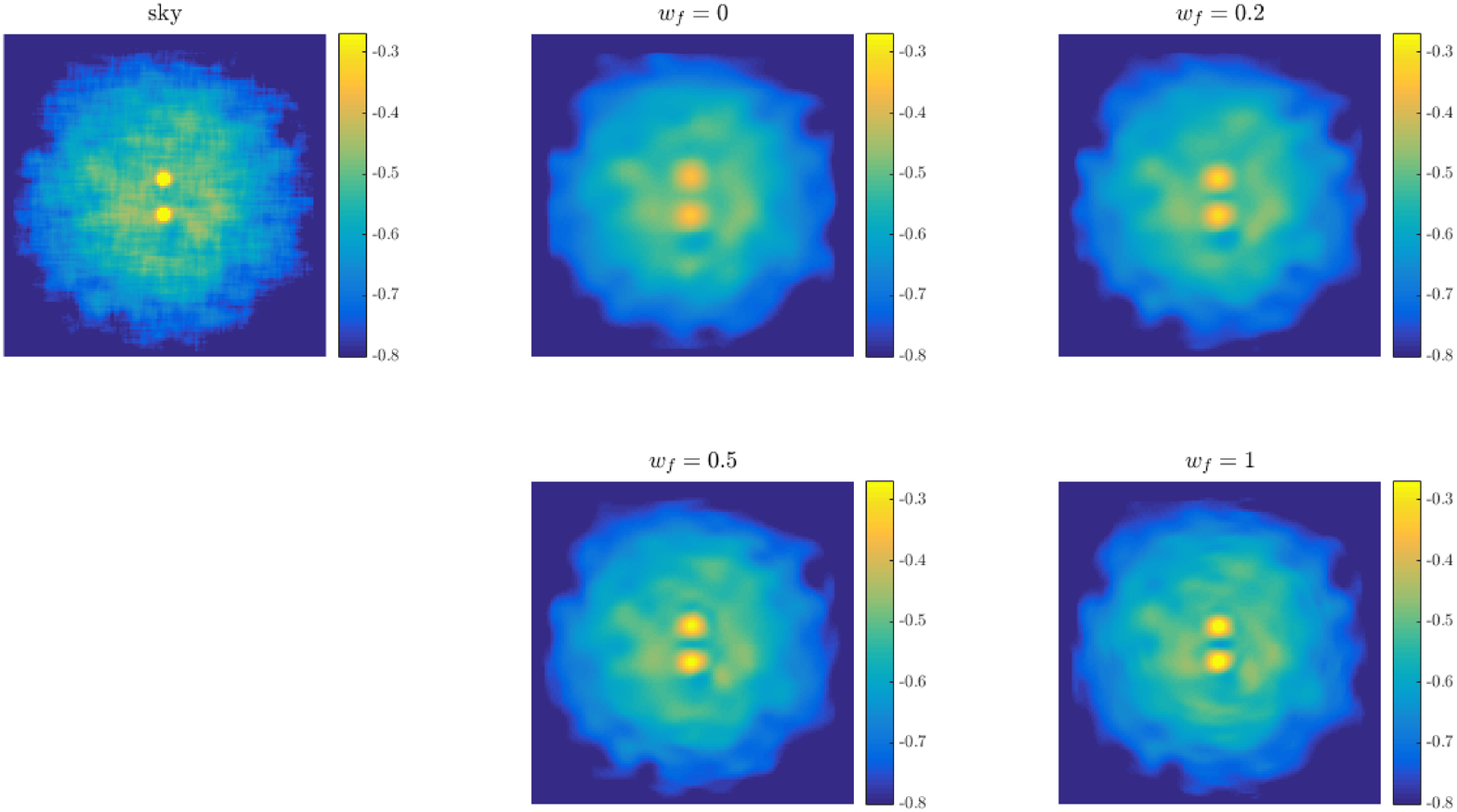}
\caption{\small Top left: ground-truth simulated image. From middle to right, reconstructed images from measurements simulated using the $uv$-coverage with $\delta{\rm t}=10$ minutes $M=3780$. From top to bottom images reconstructed from measurements with no $w$-modulation, and measurements with $w$-modulations characterised respectively by the fractions $w_f=0.2,0.5,1$.}
\label{fig_Pts_SNR40_im}
\end{minipage}
\end{figure*}

To do so, as a ground-truth image, we simulate an image of size $N=128\times 128$ consisting of two Gaussian sources with $\sigma=1.5$ pixels on a diffuse background (see Fig.~\ref{fig_Pts_SNR40_im}, top left). The diffuse background is modelled as a 2D Gaussian source with $\sigma=50$ pixels multiplied with a random Gaussian field and convolved with a highly asymmetric filter. Observations are generated using ASKAP antenna configuration for the same total observation time and direction in the sky described in Sect.~4 and with various time intervals $\delta {\rm t}=10,20,30$ minutes yielding $uv$-coverages with number of measurements $M=3780,1890,1260$  respectively, and which are denoted by settings $\rm A$, $\rm B$ and $\rm C$ respectively. Note that the three $uv$-coverages present the same bandwidth, the difference consists in the lower sampling rates for the higher time intervals.

We vary the distance separating the peaks of the two Gaussian sources of interest and adopt distances $d \in\{6,8,10,12,14\}$ pixels. To produce the band limit confusion scenario, we choose to down-sample the coverages with a factor $B_f=0.2$. Hence, in theory, the resolution obtained by such sampling is $\theta \simeq 6$ pixels at half width maximum of the PSF's primary lobe. The diffuse additive background adds complexity to resolve the two Gaussian sources since it mainly consists of low spacial frequency content, hence its energy dominates the sensed low Fourier modes. We also consider additive white Gaussian noise with ${\rm iSNR}=40$dB on the visibility data. Given this setting, we perform a set of tests while generating random $w$ values within the ranges $w_f\in\{0,0.1,0.2,0.5,1\}$. Note that, while $w_f=1$ is not a realistic setting, the objective of the study is to show the potential of the spread spectrum effect in promoting super-resolution. Measurements are simulated with the three different coverages explained above.

\begin{figure*}
\begin{minipage}{1\linewidth}
	\centering
 \includegraphics[width=0.8\textwidth, height=0.2\textwidth]{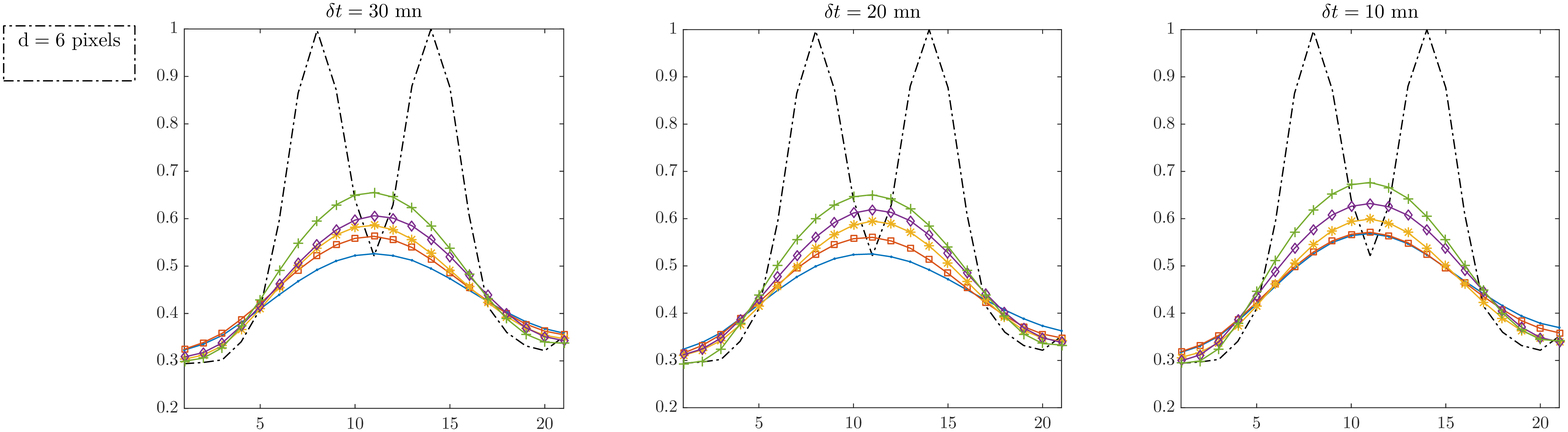}
 \includegraphics[width=0.8\textwidth, height=0.2\textwidth]{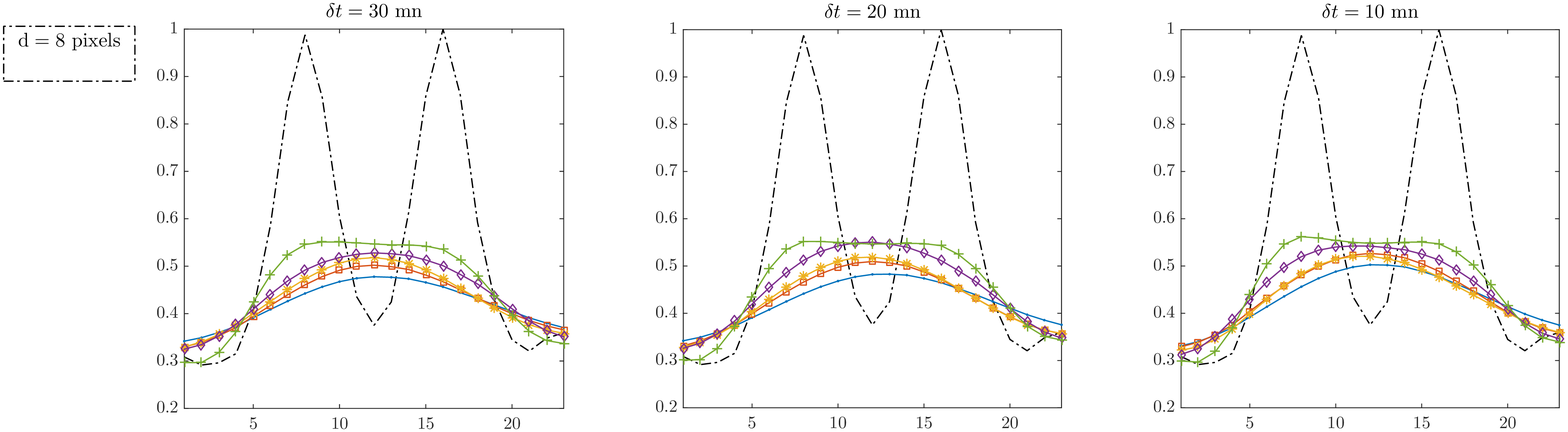}
 \includegraphics[width=0.8\textwidth, height=0.2\textwidth]{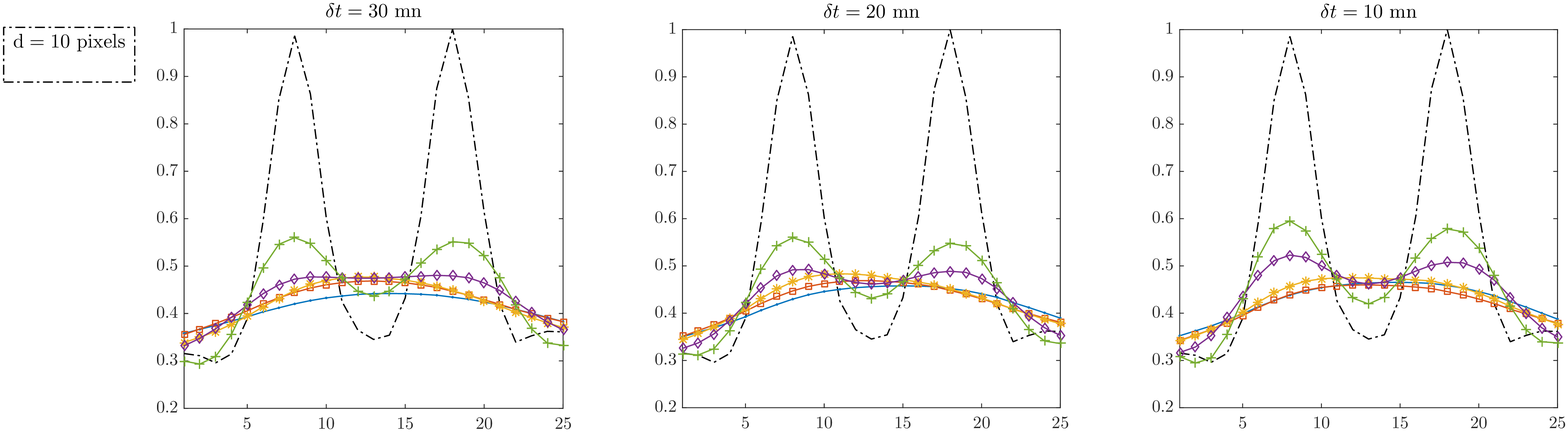}
 \includegraphics[width=0.8\textwidth, height=0.2\textwidth]{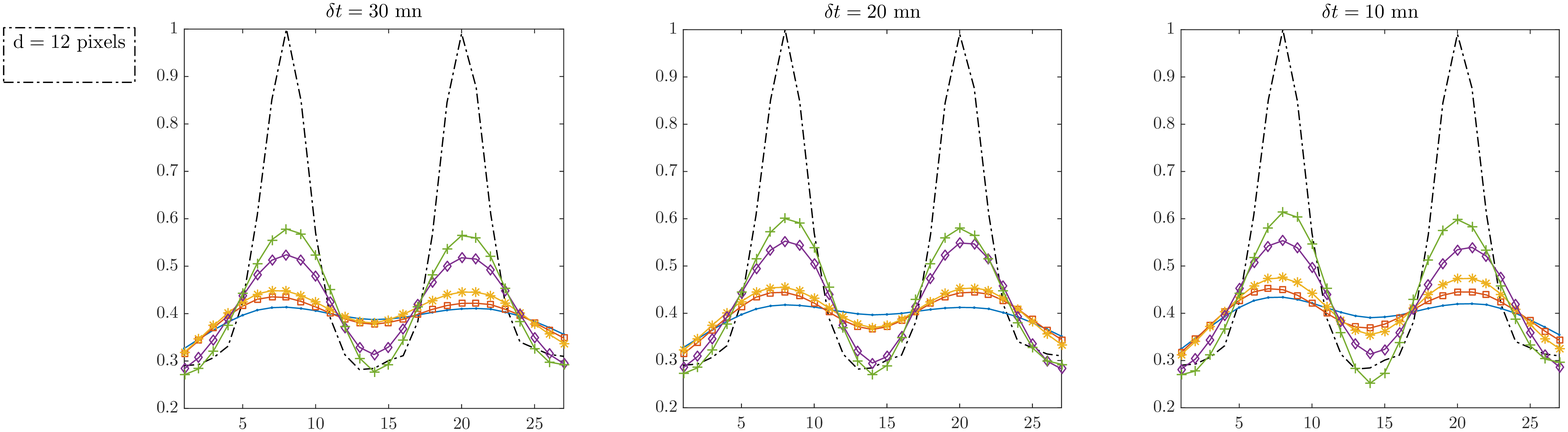}
  \includegraphics[width=0.8\textwidth, height=0.2\textwidth]{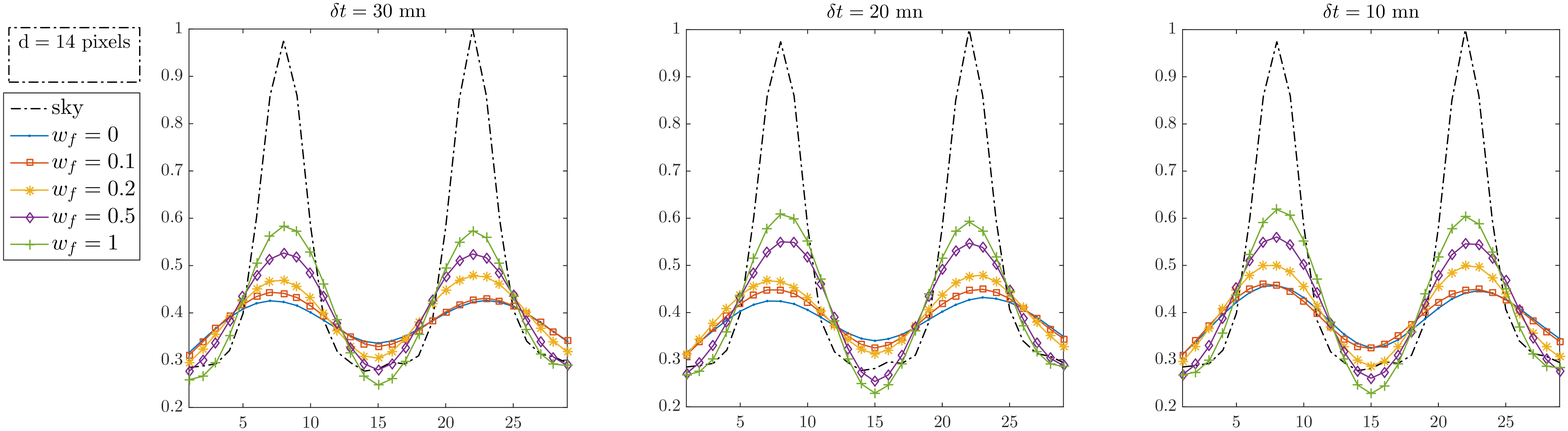}
\caption{\small Plots of the cross-section of the reconstructed images at the central column containing the peaks of the two Gaussian sources. From top to bottom, plots for the different separating distances between the two sources. From right to left, plots of the three considered settings of $uv$-coverages (A: $\delta t=10$ mn, B: $\delta t=20$ mn, C: $\delta t=30$ mn).  In each plot, is displayed the cross-section of the reconstructed images for the $w$-rates, characterised by $w_f \in \{0,0.1,0.2,0.5,1\}$.  }
\label{fig_Pts_SNR40_}

\end{minipage}
\end{figure*}

Results are shown in Fig.~\ref{fig_Pts_SNR40_}, where we display the cross-section of the reconstructed images at the central column in order to show the separability of the two Gaussian sources. For each plot, the cross-section is displayed for the three settings. In absence of $w$-modulation ($w_f =0$), for the separating distances $d=6,8,10$ pixels, the two sources are completely unresolved for all the adopted coverages. For distance $d=12$ pixels, the coverage $\rm A$ and B, having the largest number of measurements, allow for slightly resolving the two sources. For distance $d=6$ pixels, in all settings, including the presence of $w$-modulations, the two sources are unresolved, yet the higher the $w$-modulation, the more the energy coming from the two sources is concentrated.  Furthermore, we notice that for the distances $d=8,10,12$ pixels, the two sources are resolved starting from $w$-modulations with $w_f=1,0.5,0.1$ respectively. In general, the more important the $w$-rates, the better is the resolution, this is also highlighted in the reconstructed images in Fig.~\ref{fig_Pts_SNR40_im}, where estimated images are displayed for $d=12$ pixels and coverage $\rm A$, with $M=3780$ measurements. For separating distance $d=14$ pixels, the two sources are resolved in all the settings.

Increasing the number of measurements can allow for slightly better resolution as seen in the cross-section plots for $d=12$ pixels and $w=0$ (Fig.~\ref{fig_Pts_SNR40_}: fourth row); both coverages A and B with the high number of points allow to slightly resolve the two sources w.r.t the the coverage C. Yet, the $w$-modulation promotes higher resolution for relatively small $w$-rates, more significantly, as shown in the plot of the same distance with $w_f=0.1$ even with small number of measurements (coverage C). In this case, for all the adopted coverages, the two sources under scrutiny are well resolved. Generally, this test case of point source confusion limit demonstrates how powerful the $w$-term is, in terms of small-scale reconstruction as it allows to resolve sources with much higher resolution from fewer visibilities.
\section{Summary and Conclusions}
\label{sec:conclusions}
In this study, we have revisited the $w$-modulation in the context of convex optimisation, where it is incorporated in the measurement equation as a sparse thus fast measurement operator. We employ adaptive sparsification strategies on the $w$-projection operator based on the energy of its convolution kernels, and study their effects on the image reconstruction quality. Two main strategies have been investigated. The first consists in the sparsification of the Fourier-transformed chirp kernels prior to the convolution with the de-gridding kernels. The second consists in the direct sparsification of the rows of $w$-projection operator $\tilde{\bm G}$, these are the convolution kernels coupling the exact chirp kernels and the de-gridding kernels. We find that the latter strategy is prone to more sparsity of the measurement operator while preserving the image reconstruction quality. Furthermore, it is robust when no or negligible $w$-modulations are considered. This suggests that a sparser variant of the de-gridding matrix can be adopted. In general, as long as errors in the model of the visibilities introduced by the sparse approximation of the measurement operator are below the noise level, the reconstruction quality is preserved. In particular, for $\rm iSNR$ up to $\rm 40dB$, sparsification of the measurement operator (following both strategies) with energy loss fraction per row of order $\gamma= 10^{-4}$, is robust. Moreover, the sparsity ratio of the sparse measurement operator is reduced by at least a factor 2 in most cases, resulting in significantly lower memory requirements and computational time. 

Our C++ code is available online on GitHub, http://basp-group.github.io/purify/. As a future work, we plan to apply these sparsification schemes to real radio interferometric data; where we take advantage of the recent block splitting algorithmic structure proposed in \cite{onose2016}. The ability to split the data and the measurement operator into blocks and process them in a distributed manner is very promising for the applicability of $w$-projection in the context of large-scale data. Therefore, we plan to investigate efficient data block splitting strategies taking into account the $w$-terms. We also consider investigating non-constrained formulation of the radio interferometric imaging problem when adopting very sparse $w$-projection operators, yielding large errors on the backward model, possibly up to the noise level. 

We have presented a first study of a very interesting potential of the spread spectrum effect that is super-resolution. We showcased super-resolution through simulated band limited observations and point-source confusion limit experiments. Inaccessible information on the images to be recovered are resolved by adding the $w$-component into the imaging process.
We demonstrate how the $w$-term accesses Fourier modes beyond the band limit of the instrument. Important $w$-modulations allow the resolution of small-scale structures inaccessible by the band limit of the radio interferometer, in addition to an extensive enhancement of the image quality. Since the bandwidth of the $w$-modulation depends on the $w$ amplitude and the probed FoV, such potential is highly expected for radio-interferometers characterised with very long baselines and probing large field-of-views, namely the SKA. On the other hand, it is not restricted to the $w$-term. It is inherent to all DDEs, independently of their origin, thanks to their convolution nature in the imaging process. This suggests that our study is relevant for future software developments considering multiple DDEs.
\section*{Acknowledgements}
We warmly thank Federica Govoni and Matteo Murgia for providing the galaxy cluster simulated radio map analysed in this paper. This work was supported by the UK Engineering and Physical Sciences Research Council (EPSRC, grants EP/M008843/1, EP/M011089/1); the UK Science and Technology Facilities Council (STFC, grant ST/M00113X/1); the Centre for All-Sky Astrophysics (grant CE11E0090) and the gSTAR national facility at Swinburne University of Technology, funded by Swinburne and the Australian Government's Education Investment Fund.
\bibliographystyle{mnras}
\bibliography{biblio}

\begin{thebibliography}{}
\makeatletter
\relax
\def\mn@urlcharsother{\let\do\@makeother \do\$\do\&\do\#\do\^\do\_\do\%\do\~}
\def\mn@doi{\begingroup\mn@urlcharsother \@ifnextchar [ {\mn@doi@}
  {\mn@doi@[]}}
\def\mn@doi@[#1]#2{\def\@tempa{#1}\ifx\@tempa\@empty \href
  {http://dx.doi.org/#2} {doi:#2}\else \href {http://dx.doi.org/#2} {#1}\fi
  \endgroup}
\def\mn@eprint#1#2{\mn@eprint@#1:#2::\@nil}
\def\mn@eprint@arXiv#1{\href {http://arxiv.org/abs/#1} {{\tt arXiv:#1}}}
\def\mn@eprint@dblp#1{\href {http://dblp.uni-trier.de/rec/bibtex/#1.xml}
  {dblp:#1}}
\def\mn@eprint@#1:#2:#3:#4\@nil{\def\@tempa {#1}\def\@tempb {#2}\def\@tempc
  {#3}\ifx \@tempc \@empty \let \@tempc \@tempb \let \@tempb \@tempa \fi \ifx
  \@tempb \@empty \def\@tempb {arXiv}\fi \@ifundefined
  {mn@eprint@\@tempb}{\@tempb:\@tempc}{\expandafter \expandafter \csname
  mn@eprint@\@tempb\endcsname \expandafter{\@tempc}}}

\bibitem[\protect\citeauthoryear{Boyd, Parikh, Chu, Peleato  \& Eckstein}{Boyd
  et~al.}{2011}]{Boyd2011}
Boyd S.,  Parikh N.,  Chu E.,  Peleato B.,   Eckstein J.,  2011, \mn@doi
  [Found. Trends Mach. Learn.] {10.1561/2200000016}, 3, 1

\bibitem[\protect\citeauthoryear{Cand\`{e}s}{Cand\`{e}s}{2006}]{Candes2006}
Cand\`{e}s E.~J.,  2006, in Int. Congress Math.. Madrid, Spain

\bibitem[\protect\citeauthoryear{Cand\`{e}s, Wakin  \& Boyd}{Cand\`{e}s
  et~al.}{2008}]{Candes2008}
Cand\`{e}s E.~J.,  Wakin M.~B.,   Boyd S.~P.,  2008, J. Fourier Anal. Appl.,
  14, 877

\bibitem[\protect\citeauthoryear{Carrillo, McEwen  \& Wiaux}{Carrillo
  et~al.}{2012}]{Carrillo2012}
Carrillo R.~E.,  McEwen J.~D.,   Wiaux Y.,  2012, \mn@doi [\mnras]
  {10.1111/j.1365-2966.2012.21605.x}, 426, 1223

\bibitem[\protect\citeauthoryear{Carrillo, McEwen, Ville, Thiran  \&
  Wiaux}{Carrillo et~al.}{2013}]{Carrillo2013}
Carrillo R.~E.,  McEwen J.~D.,  Ville D. V.~D.,  Thiran J.-P.,   Wiaux Y.,
  2013, \mn@doi [IEEE Sig. Proc. Let.] {10.1109/LSP.2013.2259813}, 20, 591

\bibitem[\protect\citeauthoryear{Carrillo, McEwen  \& Wiaux}{Carrillo
  et~al.}{2014}]{Carrillo2014}
Carrillo R.~E.,  McEwen J.~D.,   Wiaux Y.,  2014, \mnras, 439, 3591

\bibitem[\protect\citeauthoryear{{Combettes} \& {Pesquet}}{{Combettes} \&
  {Pesquet}}{2007}]{combettes07}
{Combettes} P.~L.,  {Pesquet} J.-C.,  2007, \mn@doi [IEEE Journal of Selected
  Topics in Signal Processing] {10.1109/JSTSP.2007.910264}, \href
  {http://adsabs.harvard.edu/abs/2007ISTSP...1..564C} {1, 564}

\bibitem[\protect\citeauthoryear{{Combettes} \& {Pesquet}}{{Combettes} \&
  {Pesquet}}{2009}]{combettes09}
{Combettes} P.~L.,  {Pesquet} J.-C.,  2009, preprint, \href
  {http://adsabs.harvard.edu/abs/2009arXiv0912.3522C} {} (\mn@eprint {arXiv}
  {0912.3522})

\bibitem[\protect\citeauthoryear{{Cornwell} \& {Perley}}{{Cornwell} \&
  {Perley}}{1992}]{CornwellPerley1992}
{Cornwell} T.~J.,  {Perley} R.~A.,  1992, \aap, \href
  {http://adsabs.harvard.edu/abs/1992A%26A...261..353C} {261, 353}

\bibitem[\protect\citeauthoryear{Cornwell, Golap  \& Bhatnagar}{Cornwell
  et~al.}{2008}]{Cornwell2008}
Cornwell T.,  Golap K.,   Bhatnagar S.,  2008, \mn@doi [IEEE Selected Topics in
  Sig. Proc.] {10.1109/JSTSP.2008.2005290}, 2, 647

\bibitem[\protect\citeauthoryear{Dabbech, Mary  \& Ferrari}{Dabbech
  et~al.}{2012}]{Dabbech2012}
Dabbech A.,  Mary D.,   Ferrari C.,  2012, in 2012 IEEE International
  Conference on Acoustics, Speech and Signal Processing (ICASSP). pp
  3665--3668, \mn@doi{10.1109/ICASSP.2012.6288711}

\bibitem[\protect\citeauthoryear{{Dabbech}, {Ferrari}, {Mary}, {Slezak},
  {Smirnov}  \& {Kenyon}}{{Dabbech} et~al.}{2015}]{Dabbech2015}
{Dabbech} A.,  {Ferrari} C.,  {Mary} D.,  {Slezak} E.,  {Smirnov} O.,
  {Kenyon} J.~S.,  2015, \mn@doi [\aap] {10.1051/0004-6361/201424602}, \href
  {http://adsabs.harvard.edu/abs/2015A%26A...576A...7D} {576, A7}

\bibitem[\protect\citeauthoryear{Donoho}{Donoho}{2006}]{donoho06}
Donoho D.~L.,  2006, \mn@doi [IEEE Transactions on Information Theory]
  {10.1109/TIT.2006.871582}, 52, 1289

\bibitem[\protect\citeauthoryear{Fessler \& Sutton}{Fessler \&
  Sutton}{2003}]{Fessler2003}
Fessler J.,  Sutton B.,  2003, \mn@doi [IEEE Tran. Sig. Proc.]
  {10.1109/TSP.2002.807005}, 51, 560

\bibitem[\protect\citeauthoryear{{Garsden} et~al.,}{{Garsden}
  et~al.}{2015}]{Garsden2015}
{Garsden} H.,  et~al., 2015, \mn@doi [\aap] {10.1051/0004-6361/201424504},
  \href {http://adsabs.harvard.edu/abs/2015A%26A...575A..90G} {575, A90}

\bibitem[\protect\citeauthoryear{Greisen}{Greisen}{1998}]{greisen98}
Greisen E.,  1998, AIPS MEMORANDUM 100 National Radio Astronomy Observatory,
  Charlottesville, Virginia

\bibitem[\protect\citeauthoryear{H\"ogbom}{H\"ogbom}{1974}]{hogbom74}
H\"ogbom J.~A.,  1974, \aap, 15, 417

\bibitem[\protect\citeauthoryear{Komodakis \& Pesquet}{Komodakis \&
  Pesquet}{2015}]{Komodakis2015}
Komodakis N.,  Pesquet J.-C.,  2015, IEEE Sig. Proc. Mag., 1406.5429

\bibitem[\protect\citeauthoryear{{Li}, {Cornwell}  \& {de Hoog}}{{Li}
  et~al.}{2011}]{Li2011}
{Li} F.,  {Cornwell} T.~J.,   {de Hoog} F.,  2011, \mn@doi [\aap]
  {10.1051/0004-6361/201015045}, \href
  {http://adsabs.harvard.edu/abs/2011A%26A...528A..31L} {528, A31}

\bibitem[\protect\citeauthoryear{{McEwen} \& {Scaife}}{{McEwen} \&
  {Scaife}}{2008}]{McEwen08}
{McEwen} J.~D.,  {Scaife} A.~M.~M.,  2008, \mn@doi [\mnras]
  {10.1111/j.1365-2966.2008.13690.x}, \href
  {http://adsabs.harvard.edu/abs/2008MNRAS.389.1163M} {389, 1163}

\bibitem[\protect\citeauthoryear{{McEwen} \& {Wiaux}}{{McEwen} \&
  {Wiaux}}{2011}]{McEwen11}
{McEwen} J.~D.,  {Wiaux} Y.,  2011, \mn@doi [\mnras]
  {10.1111/j.1365-2966.2011.18217.x}, \href
  {http://adsabs.harvard.edu/abs/2011MNRAS.413.1318M} {413, 1318}

\bibitem[\protect\citeauthoryear{{Murgia}, {Govoni}, {Feretti}, {Giovannini},
  {Dallacasa}, {Fanti}, {Taylor}  \& {Dolag}}{{Murgia}
  et~al.}{2004}]{murgia2004}
{Murgia} M.,  {Govoni} F.,  {Feretti} L.,  {Giovannini} G.,  {Dallacasa} D.,
  {Fanti} R.,  {Taylor} G.~B.,   {Dolag} K.,  2004, \mn@doi [\aap]
  {10.1051/0004-6361:20040191}, \href
  {http://adsabs.harvard.edu/abs/2004A%26A...424..429M} {424, 429}

\bibitem[\protect\citeauthoryear{{Noordam} \& {Smirnov}}{{Noordam} \&
  {Smirnov}}{2010}]{Noordam2010}
{Noordam} J.~E.,  {Smirnov} O.~M.,  2010, \mn@doi [\aap]
  {10.1051/0004-6361/201015013}, \href
  {http://adsabs.harvard.edu/abs/2010A%26A...524A..61N} {524, A61}

\bibitem[\protect\citeauthoryear{{Offringa} et~al.,}{{Offringa}
  et~al.}{2014}]{Offringa2014}
{Offringa} A.~R.,  et~al., 2014, \mn@doi [\mnras] {10.1093/mnras/stu1368},
  \href {http://adsabs.harvard.edu/abs/2014MNRAS.444..606O} {444, 606}

\bibitem[\protect\citeauthoryear{Onose, Carrillo, Repetti, McEwen, Thiran,
  Pesquet  \& Wiaux}{Onose et~al.}{2016}]{onose2016}
Onose A.,  Carrillo R.~E.,  Repetti A.,  McEwen J.~D.,  Thiran J.-P.,  Pesquet
  J.-C.,   Wiaux Y.,  2016, \mn@doi [\mnras] {10.1093/mnras/stw1859}

\bibitem[\protect\citeauthoryear{{Onose}, {Dabbech}  \& {Wiaux}}{{Onose}
  et~al.}{2017}]{onose2017}
{Onose} A.,  {Dabbech} A.,   {Wiaux} Y.,  2017, preprint, \href
  {http://adsabs.harvard.edu/abs/2017arXiv170101748O} {} (\mn@eprint {arXiv}
  {1701.01748})

\bibitem[\protect\citeauthoryear{{Pratley} \& {Johnston-Hollitt}}{{Pratley} \&
  {Johnston-Hollitt}}{2016}]{PratleyJ16}
{Pratley} L.,  {Johnston-Hollitt} M.,  2016, \mn@doi [\mnras]
  {10.1093/mnras/stw1377}, \href
  {http://adsabs.harvard.edu/abs/2016MNRAS.462.3483P} {462, 3483}

\bibitem[\protect\citeauthoryear{{Pratley}, {McEwen}, {d'Avezac}, {Carrillo},
  {Onose}  \& {Wiaux}}{{Pratley} et~al.}{2016}]{pratley16A}
{Pratley} L.,  {McEwen} J.~D.,  {d'Avezac} M.,  {Carrillo} R.~E.,  {Onose} A.,
   {Wiaux} Y.,  2016, preprint, \href
  {http://adsabs.harvard.edu/abs/2016arXiv161002400P} {} (\mn@eprint {arXiv}
  {1610.02400})

\bibitem[\protect\citeauthoryear{{Sault}, {Teuben}  \& {Wright}}{{Sault}
  et~al.}{1995}]{Sault1995}
{Sault} R.~J.,  {Teuben} P.~J.,   {Wright} M.~C.~H.,  1995, in {Shaw} R.~A.,
  {Payne} H.~E.,   {Hayes} J.~J.~E.,  eds,  Astronomical Society of the Pacific
  Conference Series Vol. 77, Astronomical Data Analysis Software and Systems
  IV. p.~433 (\mn@eprint {} {astro-ph/0612759})

\bibitem[\protect\citeauthoryear{{Schwab}}{{Schwab}}{1984}]{Schwab1984}
{Schwab} F.~R.,  1984, in {Roberts} J.~A.,  ed., Indirect Imaging. Measurement
  and Processing for Indirect Imaging. pp 333--346

\bibitem[\protect\citeauthoryear{Setzer, Steidl  \& Teuber}{Setzer
  et~al.}{2010}]{Setzer2010}
Setzer S.,  Steidl G.,   Teuber T.,  2010, \mn@doi [J. Vis. Comun. Image
  Represent.] {10.1016/j.jvcir.2009.10.006}, 21, 193

\bibitem[\protect\citeauthoryear{{Wakker} \& {Schwarz}}{{Wakker} \&
  {Schwarz}}{1988}]{Wakker88}
{Wakker} B.~P.,  {Schwarz} U.~J.,  1988, \aap, \href
  {http://adsabs.harvard.edu/abs/1988A%26A...200..312W} {200, 312}

\bibitem[\protect\citeauthoryear{Wenger, Magnor, Pihlstr\"{o}sm, Bhatnagar  \&
  Rau}{Wenger et~al.}{2010}]{wenger10}
Wenger S.,  Magnor M.,  Pihlstr\"{o}sm Y.,  Bhatnagar S.,   Rau U.,  2010,
  PASP, 122, 1367

\bibitem[\protect\citeauthoryear{Wiaux, Jacques, Puy, Scaife  \&
  Vandergheynst}{Wiaux et~al.}{2009a}]{Wiaux09b}
Wiaux Y.,  Jacques L.,  Puy G.,  Scaife A. M.~M.,   Vandergheynst P.,  2009a,
  \mn@doi [\mnras] {10.1111/j.1365-2966.2009.14665.x}, 395, 1733

\bibitem[\protect\citeauthoryear{Wiaux, Puy, Boursier  \& Vandergheynst}{Wiaux
  et~al.}{2009b}]{Wiaux09}
Wiaux Y.,  Puy G.,  Boursier Y.,   Vandergheynst P.,  2009b, \mn@doi [\mnras]
  {10.1111/j.1365-2966.2009.15519.x}, 400, 1029

\bibitem[\protect\citeauthoryear{Wolz, McEwen, Abdalla, Carrillo  \&
  Wiaux}{Wolz et~al.}{2013}]{Wolz2013}
Wolz L.,  McEwen J.~D.,  Abdalla F.~B.,  Carrillo R.~E.,   Wiaux Y.,  2013,
  \mn@doi [\mnras] {10.1093/mnras/stt1707}, 436, 1993

\makeatother
\end{thebibliography}
\bsp	
\label{lastpage}
\end{document}